\documentclass[aps,pre,reprint,superscriptaddress,amsmath,amssymb,showpacs,showkeys]{revtex4-1}

\usepackage{graphicx}\graphicspath{{images/}}
\usepackage{subfigure}
\usepackage{dcolumn}
\usepackage{float}
\usepackage{bm}
\usepackage[colorlinks,linkcolor=blue,citecolor=blue,urlcolor=blue]{hyperref}
\usepackage{multirow}
\newcommand{\figref}[1]{Fig.~\ref{#1}}
\newcommand{\equref}[1]{Eq.~(\ref{#1})}
\newcommand{\tabref}[1]{Table~\ref{#1}}
\newcommand{\secref}[1]{Sec.~\ref{#1}}
\usepackage[utf8]{inputenc}
\usepackage[T1]{fontenc}
\usepackage{mathptmx}

\newcommand{\pos}[2]{(#1,#2)}

\begin{document}

\title{Koopman Operator and Phase Space Partition of Chaotic Maps}

\author{Cong Zhang}
\affiliation{School of Science, Beijing University of Posts and Telecommunications, Beijing 100876, China}
\author{Yueheng Lan}
\email{lanyh@bupt.edu.cn}
\affiliation{School of Science, Beijing University of Posts and Telecommunications, Beijing 100876, China}
\affiliation{State Key Lab of Information Photonics and Optical Communications, Beijing University of Posts and Telecommunications, Beijing 100876, China}
\date{\today}

\begin{abstract}
Koopman operator describes evolution of observables in the phase space, which could be used to extract characteristic dynamical features of a nonlinear system. Here, we show that it is possible to carry out interesting symbolic partitions based on properly constructed eigenfunctions of the operator for chaotic maps. The partition boundaries are the extrema of these eigenfunctions, the accuracy of which is improved by including more basis functions in the numerical computation. The validity of this scheme is demonstrated in well-known 1-d and 2-d maps. It seems no obstacle to extend the computation to nonlinear systems of high dimensions, which provides a possible way of dissecting complex dynamics. 
\end{abstract}
\maketitle

\section{Introduction\label{sec:intro}}

For complex systems,  due to the lack of understanding of underlying physical principles or generic complexity, it is impossible to make an effective description or even pin down the major characteristics of its dynamics. With the 
rapid development of data collecting and storing technology, however, it is indeed feasible to obtain large amount of data through large-scale simulation, detailed experiments or high-resolution observation.    Although these data may be handled with modern intelligent "black box", {\em e.g.}, various neural networks~\cite{hassoun1995fundamentals}, a deeper understanding of key dynamical features of the system remains desperately desirable and thus of essential importance. 

Dynamical systems theory often studies orbit structures in phase space and their changes under parameter variations. In linear or integrable systems, all the orbits are arranged regularly and qualitative features of the system may be estimated with very high accuracy. However, when the dynamics becomes chaotic, which is the case for most nonlinear systems in high dimensions, traditional analytic approaches often do not apply~\cite{strogatz2001nonlinear}. Even numerical computation becomes unreliable~\cite{holmes1996coherent} in these systems, where the orbit structure becomes extremely complicated and sensitive to initial condition or external perturbation. It seems necessay to introduce a statistical assumption in this context like what has been done in statistical mechanics~\cite{landau1973statistical}. However, the lack of a variational principle and non-stationarity in general requires that it should start from first principles. Otherwise important dynamical features may be missed in the treatment of nonlinear systems far from equilibrium. This concern has been raised long time ago in an effort to build rigorous framework of statistical physics, where Liouville equation is viewed as a starting point by many~\cite{prigogine2017non}. In this type of approach, the focus is shifted from individual orbits to the evolution of specific functions defined in the phase space. In Hamiltonian systems, this job is done by the well-known Liouville operator while in general dynamical systems, Koopman operator is the analogue. 

The Koopman operator was put forward by B. O. Koopman in 1931~\cite{koopman1931hamiltonian}, which  describes evolution of functions defined in the phase space of a dynamical system and is closely related to Liouville operator~\cite{gaspard1995spectral,gaspard2001liouvillian}. The eigenvalues and eigenfunctions of this linear operator capture global features of system dynamics~\cite{budivsic2012applied}, which were first used by Mezi{\' c} {\em et al} for ergodic partitions in systems with heterogeneous dynamics~\cite{mezic2004comparison}, and for the possibility of simplifying high-dimensional dynamical systems~\cite{mezic2005spectral}. In recent years, the dynamical mode decomposition (DMD) algorithm based on Koopman operator and its improved versions have been extensively tested on different occasions including the power system~\cite{susuki2011nonlinear,susuki2012nonlinear}, building energy efficiency~\cite{eisenhower2010decomposing,georgescu2012creating} and fluid systems~\cite{schmid2011applications,bagheri2013koopman,mezic2013analysis}. The robustness of the DMD algorithm is discussed in \cite{schmid2010dynamic} and the fine structure of the spectrum could in principle be computed with rigorous convergence guarantee from measured data~\cite{korda2020data}. In practice, Hankel type of matrices is easy to construct for a convenient computation of the spectra by properly arranging the data series~\cite{arbabi2017ergodic}. Further exploration reveals a connection of the linearizing transformation to the spectrum of Koopman operator~\cite{lan2013linearization}, which provides a link to 
another powerful tool in the study of nonlinear dynamics - the symbolic dynamics.  

Symbolic dynamics originates from the abstract topological theory of dynamical systems~\cite{morse1938symbolic}, and is gradually developed and perfected in the study of one-dimensional maps \cite{crutchfield1982symbolic}. In symbolic dynamics, the state is expressed as an infinite sequence of finite set of symbols and the dynamics is represented by symbol shifts, although certain metric information is lost in this translation~\cite{robinson1998dynamical}. To establish symbolic dynamics in a nonlinear system, one of the most critical problems is how to find a reasonable symbolic partition~\cite{hao1991symbolic}. In each partitioned area, there should be no folding which is typical in the phase space of a chaotic mapping, where some monotonicity or linearity could be envisioned. If the unstable direction is one-dimensional, then the partition is realized by a set of "boundary points", determined usually by checking the intersections of stable and unstable manifolds~\cite{biham1989characterization,jaeger1997structure,grassberger1985generating}. Interestingly, we find in this paper that
properly selected eigenfunctions of the Koopman operator may be used to divide the phase space in a way that is  consistent with the symbolic partition. This observation is successfully confirmed in several 1-d maps with different characteristics as well as in the well-known H{\'e}non map when it is chaotic.

The rest of the paper is organized as follows. The concept and some properties of the Koopman operator are given in \secref{sec:koopman}, together with a discussion of phase space partition based on its eigenfunctions. Several chaotic maps in one or two dimensional phase space are used as examples to demonstrate the implementation of the theory in \secref{sec:sample}. The correspondence between the extrema of eigenfunctions and the boundary points of symbolic partition is checked in great detail. In \secref{sec:conclusion}, the whole paper is summarized and possible future directions are pointed out. 

\section{Koopman operator and partition of the phase space\label{sec:koopman}}
\subsection{Koopman Operator and Dynamical Mode}
A discrete-time dynamical system, for any point $x_p\in \mathbf{P}$ in the phase space $\mathbf{P}$, defines $x_{p+1}=T(x_p)$, where $p$ is the number of iterations, and $T$ gives the iteration law. The Koopman operator $U$ acts on a phase space function $f(x)$ giving
\begin{equation}
Uf(x)=f(T(x))
\end{equation}
which is a linear operator that can be spectrally decomposed. The eigenvalue $\lambda$ and the eigenfunction $\phi(x)$ of the Koopman operator satisfies 
\begin{equation}
U\phi(x)=\phi(T(x))=\lambda\phi(x)
\,.
\label{eq:Koopman_eigen}
\end{equation}

For example, a one-dimensional discrete dynamical system $x_{p+1}=2x_p$ gives $T(x)=2x$. For an observable function $f(x)=x^2$, $Uf(x)=f(T(x))=f(2x)=4x^2=\tilde{f}(x)$. That is, the Koopman operator acts on the function and induces the following mappings
\begin{subequations}
	\begin{align}
	x^2 &\stackrel{U}{\longrightarrow} 4x^2\\
	f(x) &\stackrel{U}{\longrightarrow} \tilde{f}(x)
\,.
	\end{align}
\label{eq:Koopman_example}%
\end{subequations}
According to \equref{eq:Koopman_eigen}, we may obtain two eigenfunctions $\phi_1(x)=1$ and $\phi_2=x$, and their corresponding eigenvalues are $\lambda_1=1$ and $\lambda_2=2$.

In dynamical systems, the eigenvalues and eigenfunctions of Koopman operator is bestowed with specific characters of system dynamics. According to the definition of the Koopman operator, we have for the eigenfunction $\phi(x)$
\begin{equation}
\begin{aligned}
\phi(x_p)&=U\phi(x_{p-1})=\lambda\phi(x_{p-1})\\
&=\lambda U\phi(x_{p-2})=\lambda^2\phi(x_{p-2})\\
&\cdots\\
&=\lambda^{n-1}U\phi(x_0)=\lambda^n\phi(x_0)
\end{aligned}
\,.
\label{eq:koop_phi}
\end{equation}
We see that the eigenfunctions of the Koopman operator are related by integer powers of eigenvalues along an orbit. Therefore, these functions are very special in the study of phase space dynamics of nonlinear systems, which may be called nonlinear modes  in analogy with the  eigen-modes in a linear  system. If enough modes are identified, the global dynamics should become clear. However, which are the important modes and how to find them, what kind of roles they play are all important questions in the Koopman analysis. 

In particular, when $\lambda=1$, the eigenfunction $\phi(x_p)=\phi(x_{p-1})=\cdots=\phi(x_{0})$($p\in \mathbb{Z}$ is the discrete time index), so the eigenfunction does not change on an orbit, i.e., $\phi(x)=C$($C$ is a constant). Therefore, different constant eigenfunctions indicate different classes of solutions. When $|\lambda|=1$ (i.e., $\lambda=e^{i\theta}$), the eigenfunction $|\phi(x_p)|=|\phi(x_{p-1})|=\cdots=|\phi(x_{0})|$, the modulus of which does not change with time (while the phase changes), and thus corresponds to a constant of motion while the phase increases linearly and leads to oscillations along an orbit.

For a complex dynamical system, it is often difficult to describe an orbit in the whole phase space. If we divide the phase space of the dynamical system into distinct regions resulting from different dynamical modes, system evolution may be expressed in terms of itinerary of visiting these regions, that is, it is possible to carry out the dynamical mode decomposition (DMD) of the phase space \cite{alla2017nonlinear}. The eigenvalues and eigenfunctions of the Koopman operator provide us with a way for such division. For eigenfunctions with eigenvalues of $|\lambda|=1$, the points with the same modulus belong to an invariant set. The work in \cite{brunton2016koopman} explores finite-dimensional linear representations of nonlinear dynamical systems by restricting the Koopman operator to an invariant set. If the phase space can be divided with these invariant sets, we can realize a block description of dynamical modes in the phase space.

The idea of describing the phase space with qualitatively different regions reminds us of symbolic dynamics. In a discrete dynamical system, we can name different regions of the phase space with different symbols, for example, using the symbol "0" and "1" to divide the phase space into two different regions. Each point corresponds to an infinite sequence of symbol "0" or "1" according to its visitation of these two regions during map iterations. For a permissible finite sequence, a group of points may be identified which have this sequence as a common visitation itinerary at the initial finite steps. In this way, we are able to qualitatively determine how many different orbits are generated in an evolution and estimate the importance of individual orbits~\cite{cvitanovic2020chaos}. Even when traditional analytical calculations do not apply, we may still try to invoke a coarse-grained symbolic dynamics to study properties of dynamical systems.

We call the critical points that divide the phase space as "boundary points". In symbolic dynamics, one of the basic issues is how to find "boundary points". As discussed above, specific eigenfunctions of the Koopman operator may help distinguish regions with different characteristics, which could in principle be used to locate these boundary points for a symbolic partition.

\subsection{Numerical representation}

We obtain the eigenvalues and eigenfunctions of the Koopman operator by numerical calculation. In order to accurately describe an eigenfunction, we can select a group of basis functions $g_1(x),g_2(x),\cdots,g_m(x)$, which make up a finite-dimensional function space, For any function $f(x)$, we may approximate it with the expansion
\begin{equation}
f(x)=\sum_{i=1}^m\alpha_ig_i(x)
\,.
\end{equation}
Of course, the accuracy of the approximation depends on $f(x)$ itself as well as the truncation order $m$. If we can describe the evolution of all basis functions in this function space, an approximation of the Koopman operator is obtained.

More explicitly, we select the data at the time point $p$ as $\{x_{p_1},x_{p_2},\cdots,x_{p_n}\}$, and when the time moves to $p+1$, the data evolves to $\{x_{p_1+1},x_{p_2+1},\cdots,x_{p_n+1}\}$. The basis function ${g_i(x)}$,$i=1,2,\cdots,m$, as well as the evolved functions are expressed as column vectors at the known data points $\{x_{p_1},x_{p_2},\cdots,x_{p_n}\}$, thus forming two data matrices of dimension $n\times m$:
\begin{subequations}
	\begin{align}
	&\begin{aligned}
		K&=\left(g_1(x_p),g_2(x_p),\cdots,g_m(x_p)\right)\\
		&=\begin{pmatrix}
		g_1(x_{p_1})&g_2(x_{p_1})&\cdots&g_m(x_{p_1})\\
		g_1(x_{p_2})&g_2(x_{p_2})&\cdots&g_m(x_{p_2})\\
		\vdots&\vdots&\ddots&\vdots\\
		g_1(x_{p_n})&g_2(x_{p_n})&\cdots&g_m(x_{p_n})
		\end{pmatrix}
	\end{aligned}
\,,
\label{eq:koop_K}\\
	&\begin{aligned}
		L&=\left(g_1(x_{p+1}),g_2(x_{p+1}),\cdots,g_m(x_{p+1})\right)\\
		&=\begin{pmatrix}
		g_1(x_{p_1+1})&g_2(x_{p_1+1})&\cdots&g_m(x_{p_1+1})\\
		g_1(x_{p_2+1})&g_2(x_{p_2+1})&\cdots&g_m(x_{p_2+1})\\
		\vdots&\vdots&\ddots&\vdots\\
		g_1(x_{p_n+1})&g_2(x_{p_n+1})&\cdots&g_m(x_{p_n+1})
		\end{pmatrix}\\
		&=\left(\tilde{g}_1(x_p),\tilde{g}_2(x_p),\cdots,\tilde{g}_m(x_p)\right)
	\end{aligned}\label{eq:koop_L}
\,.
	\end{align}
	\label{eq:koop_KL}%
\end{subequations}
Each column of $K$ and $L$ is a discrete approximation of a function in the phase space. When the number of points is large enough, any smooth function with mild oscillation could be represented with reasonable accuracy. $K$ and $L$ are related by the Koopman operator
\begin{equation}
K\stackrel{U}{\longrightarrow}L
\,.
\label{eq:koop_U1}
\end{equation}
If $L$ is also viewed as a function of $x_p$, the matrix representation of the Koopman operator can be determined by \equref{eq:koop_U1}. More explicitly, we choose to write the relation as
\begin{equation}
K\tilde{U}=L
\,.
\label{eq:koop_U2}
\end{equation}
After obtaining the matrix representation $\tilde{U}$ of the Koopman operator, we can further compute the eigenvalues and eigenfunctions.

The Koopman operator acts in an infinite dimensional function space. Although our function space dimension is only m, it has been proved~\cite{govindarajan2019approximation} that both the spectra measures and projectors of the operators converge to their infinite-dimensional counterparts in the limit $m \to \infty$.

After selecting appropriate basis functions and constructing the data matrices $K$ and $L$, we obtain $m$ eigenvalues and eigenfunctions of the Koopman operator based on \equref{eq:koop_U2}. According to \equref{eq:koop_phi}, we are concerned with the eigenvalue $\lambda=1$, because the corresponding eigenfunction does not change along an orbit and thus may be used to characterize different ergodic components. However, our calculation is done on a finite projection of the infinite dimensional space, which brings errors inevitably, so in practice the eigenvalue of $\lambda\approx1$ and its corresponding eigenfunctions are considered.

\subsection{Function Space: Selection of Basis Functions}
According to the discussion in the previous section, to construct the data matrix $K$ and $L$, we have to select a group of basis functions $\{g_i(x)\},i=1,2,\cdots,m$, which define the function space for the approximation and should capture the main dynamical features of the system under investigation. A proper selection will make the computation much easier and allows a robust reconstruction of relevant eigenfunctions, thus constituting an essential part of the whole scheme.

For computational efficiency and convenience, usually a complete set of orthogonal functions are selected. As usual, the orthogonality of the basis functions $g_i(x)$ is given by
\begin{equation}
\langle g_i, g_j\rangle= \delta_{ij}
\,,
\end{equation}
where the inner product $\langle f,h \rangle=\int_{\mathcal{M}}{f(x)h(x)}dx$ is defined for any two functions $f(x)$ and $g(x)$. In a rectangular area, depending on the boundary conditions, exponential functions or Legendre polynomials are commonly used orthogonal basis functions. However, both of them are non-local functions, which seem hard to capture functions defined on a submanifold in the phase space. For a dissipative evolution, the asymptotic attractor often concentrates on a low-dimensional hyper-surface, and a lot of these global functions are needed to depict this dimension reduction. A more convenient representation is given by sets of local functions whose supports are bounded.

The following are some examples of basis functions in an interval on the $x-$axis
\begin{subequations}
	\begin{align}
	g_R(x)&=
	\begin{cases}
	\sqrt{m},\ &(\dfrac{i-1}{m}\leqslant x<\dfrac{i}{m})\\
	0,\ &(otherwize)
	\end{cases},\ i=1,2,\cdots,m\label{eq:basis_rect}\\
	g_G(x)&=Cexp\left(-\dfrac{(x-x_i)^2}{2d_j^2}\right), \ x_i=\frac{i}{m}-\frac{1}{2m}\label{eq:basis_gauss}\\
	g_F(x)&=e^{ik(2\pi)x},\ k=-m,-(m-1),\cdots,m-1,m\label{eq:basis_four}\\
	g_L(x)&=\sqrt{\dfrac{2k+1}{2}}P_i(x),\ k=0,1,\cdots,m\label{eq:basis_legen}
	\end{align}
\label{eq:basis_function}%
\end{subequations}
where a finite resolution is assumed and signalled by the finite range of the running indices. \equref{eq:basis_rect} is a basis of rectangular functions defined in the interval $[0,1]$, which is orthogonal and local, while the Gaussian basis \equref{eq:basis_gauss} is nearly local and approximately orthogonal, the balance of which is controlled by the width $d_j$ and the center $x_i=\frac{i}{m}-\frac{1}{2m}$. For later reference, the Fourier basis \equref{eq:basis_four} in $[0,1]$ and the Legendre polynomials \equref{eq:basis_legen} in $[-1,1]$ are also displayed, which are orthogonal but non-local.

Nevertheless, for high-dimensional systems, all the above basis may not be good since the attractor which for finite resolution may be approximated by an intricately curved hyper-surface needs a lot of these functions to characterize, the number being exponentially increasing with the dimension of the phase space. On the other hand, a heterogeneous distribution of the orbits on the attractor may as well entail a slow convergence of function expansion and hence requires a large set of basis functions. One solution to both problems is the utilization of the natural basis constructed from the evolution data itself. We may take $n$ points along an orbit to approximate a function defined in the relevant region of the phase space. For example, the $n$ values of the observable $x$, {\em i.e.}, $g_i(x)=(x_i,x_{i+1},\cdots,x_{n+i-1})^T$ defines a discrete representation of a linear function that evolves with time marked with the index $i$, according to which Kutz {\em et al}~\cite{brunton2017chaos} proposed an alternative view of Koopman (HAVOK) analysis. They put such defined functions side by side to form a Hankel matrix as the basis set of functions $\{g_i(x)\}\,, i=1,\cdots, m $, whose one-step evolution is obtained by applying the Koopman operator to $g_i(x)$:
\begin{equation}
\begin{aligned}
Ug_i(x)&=U(x_i,x_{i+1},\cdots,x_{n+i-1})^T\\
&=(x_{i+1},x_{i+2},\cdots,x_{n+i})^T\\
&=g_{i+1}(x)
\end{aligned}
\label{eq:koop_U3}
\,,
\end{equation}
which then provides the data matrix $K$ and $L$ in \equref{eq:koop_KL}. As in \equref{eq:koop_U1}, the matrix approximation of the Koopman operator is thus obtained. All the basis functions are chosen from 
the given evolution data, which are mainly supported on the relevant part of the phase space and already contains intrinsic properties of the nonlinear system. Nevertheless, if transient dynamics is also of our concern, the basis functions listed in \equref{eq:basis_function} should be used, which usually involves stable manifolds in the phase space, as shown in the 2-d example below.

\section{Application to several typical examples\label{sec:sample}}
The scheme of previous section will be applied to several typical chaotic maps including variants of the tent map, the logistic map in one dimension and the H\'{e}non map in two-dimensional phase space. The eigenvalues and eigenfunctions of the Koopman operator are calculated using discrete evolution data, which may be checked and used for the partition of the phase space. 

\subsection{One-dimensional map: the tent and the logistic map}
The tent map is a one-dimensional piecewise linear map defined on interval $[0,1]$, named after its functional image resembling a tent 
\begin{equation}
x_{n+1}=f(x_n)=1-2\left|x-\frac{1}{2}\right|=
\begin{cases}
\begin{aligned}
2x_n&,\ x\in [0,\frac{1}{2})\\
2-2x_n&,\ x\in [\frac{1}{2},1]
\end{aligned}
\end{cases}
\label{eq:tent}
\end{equation}
which has two fixed points: $x_1^*=0$ and $x_2^*=\frac{2}{3}$. The tent map is a chaotic map with a uniform invariant measure, which stretches all local segments uniformly and then folds in a symmetric way. On the other hand, the
logistic map 
\begin{equation}
x_{n+1}=f(x_n)=\gamma x_n(1-x_n),\ x_n\in [0,1]
\label{eq:logistic}
\end{equation}
stretches non-uniformly and the invariant measure is singular. It has two fixed points: $x_1^*=0$ and $x_2^*=1-\frac{1}{\gamma}$ as well. In the following analysis, we take $\gamma=4$ and the system is in a chaotic state.

\subsubsection{The Eigenfunctions of Koopman Operator and the Partition of Phase Space}

\begin{figure*}
	\begin{minipage}{0.48\linewidth}
		\centerline{\includegraphics[width=8.5cm]{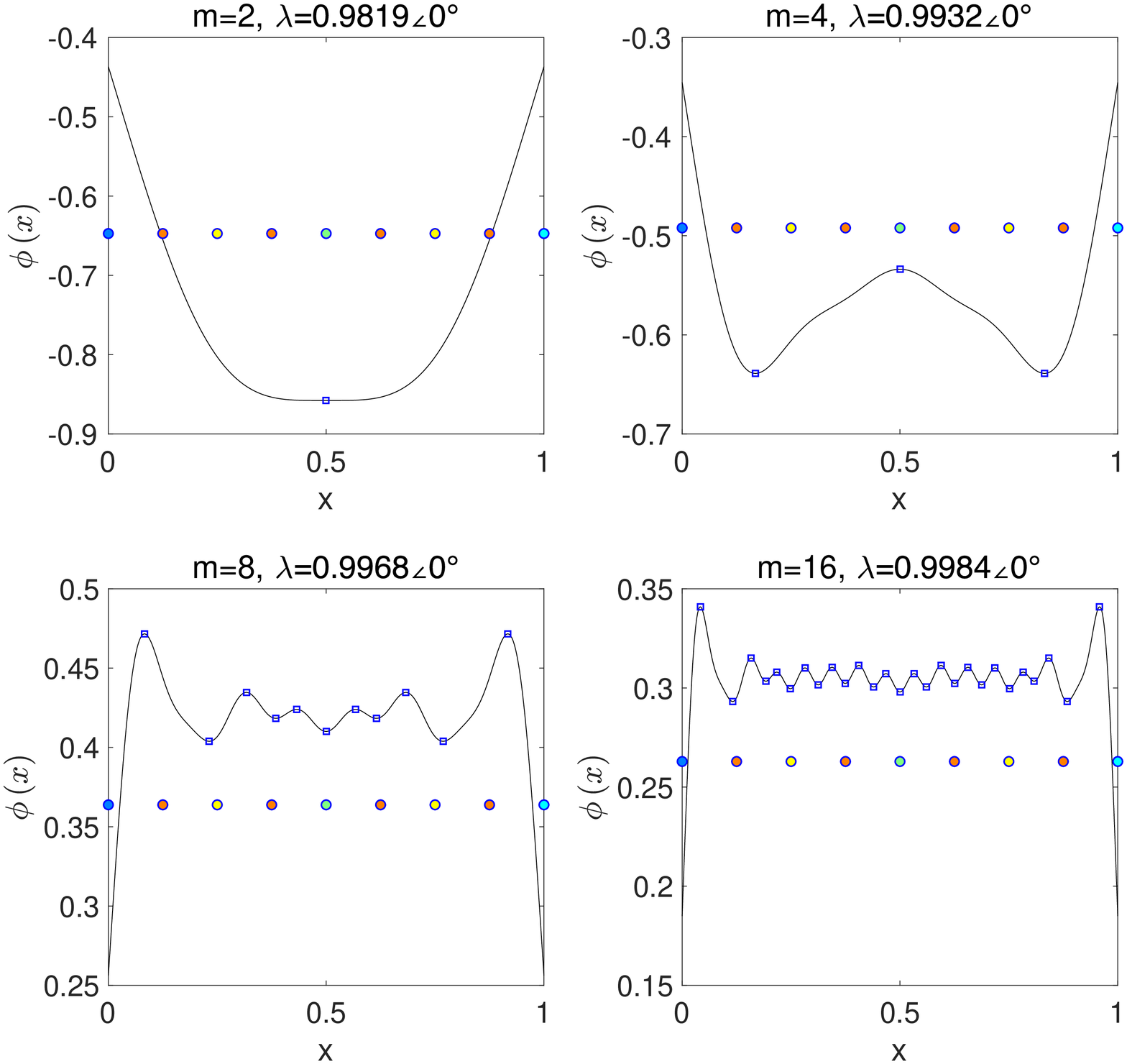}}
		\centerline{(a) tent map \equref{eq:tent}}
	\end{minipage}
	\hfill
	\begin{minipage}{0.48\linewidth}
		\centerline{\includegraphics[width=8.5cm]{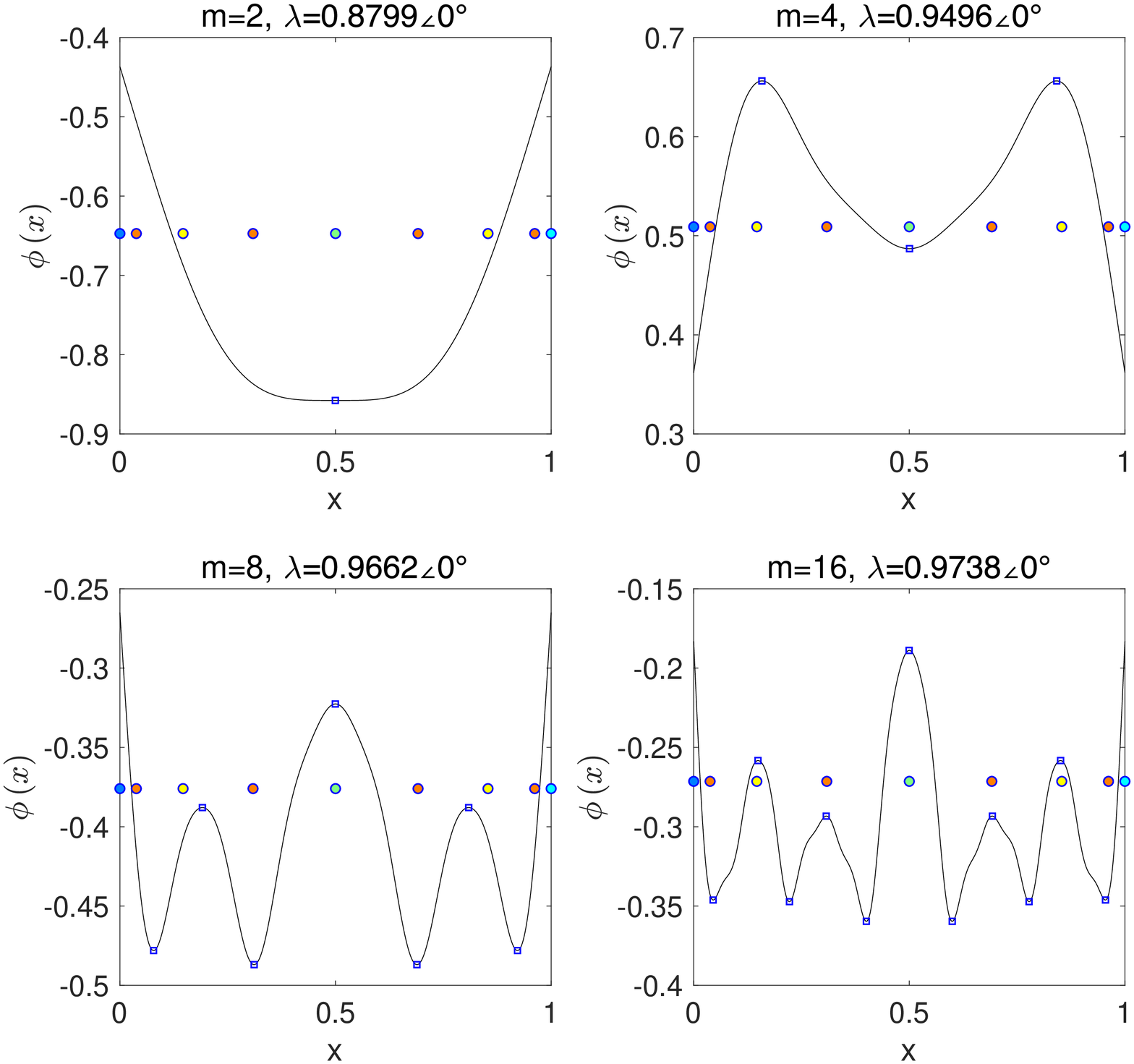}}
		\centerline{(b) logistic map \equref{eq:logistic}}
	\end{minipage}
	\caption{Comparison of extremum points (empty squares) and boundary points (colored points, indicating different levels): eigenvalues $\lambda$ (with moduli closest to 1, expressed in modulus and angle) and eigenfunctions (black lines) of the Koopman operator ($n=1000, m=2,4,8,16$) computed with the Gaussian basis for (a) the tent map \equref{eq:tent} and (b) the logistic map \equref{eq:logistic}.\label{fig:Koopman_eigen_Gauss_boundary_n1000m50}}
\end{figure*}

In the tent and logistic map, we employ the Gaussian basis and take $n=1000$, and the number of basis functions $m=2,4,8,16$. Usually we get many eigenvalues and eigenfunctions, the number of which depends on $m$. As we mentioned in the previous section, we are more concerned with the eigenvalues close to 1 and the corresponding eigenfunctions. We plot the selected eigenfunctions (unless otherwise specified, when the eigenfunction is complex, we use the real part) in \figref{fig:Koopman_eigen_Gauss_boundary_n1000m50}.

In \figref{fig:Koopman_eigen_Gauss_boundary_n1000m50}, we choose the eigenvalue of the modulus closest to 1. Each eigenfunction is composed of several wave packets. When $m$ increases, the dimension of our function space also rises, and the approximation to the Koopman operator gains accuracy. 

\begin{figure*}
	\begin{minipage}{0.48\linewidth}
		\centerline{\includegraphics[width=8.5cm]{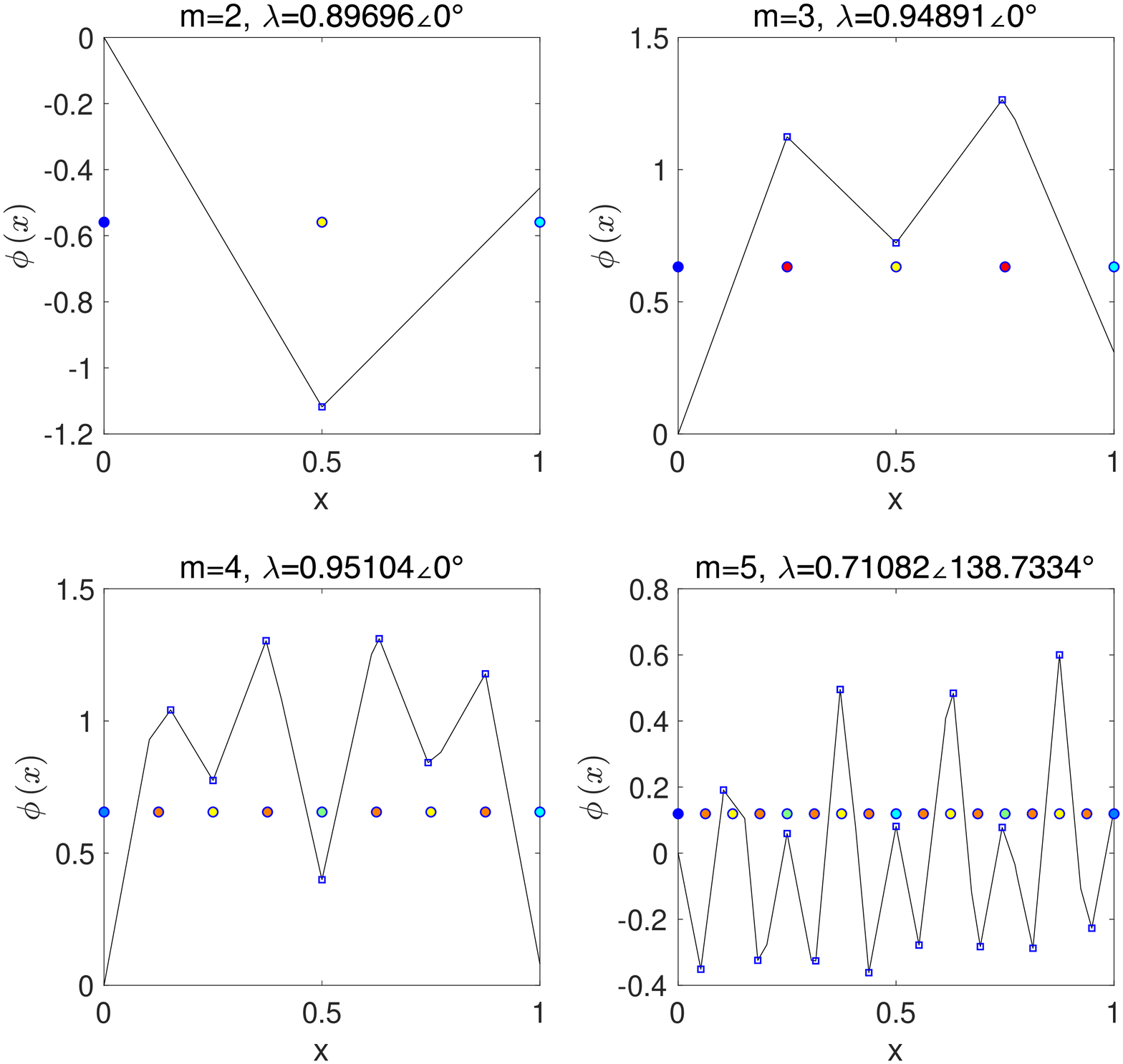}}
		\centerline{(a) tent map \equref{eq:tent}}
	\end{minipage}
	\hfill
	\begin{minipage}{0.48\linewidth}
		\centerline{\includegraphics[width=8.5cm]{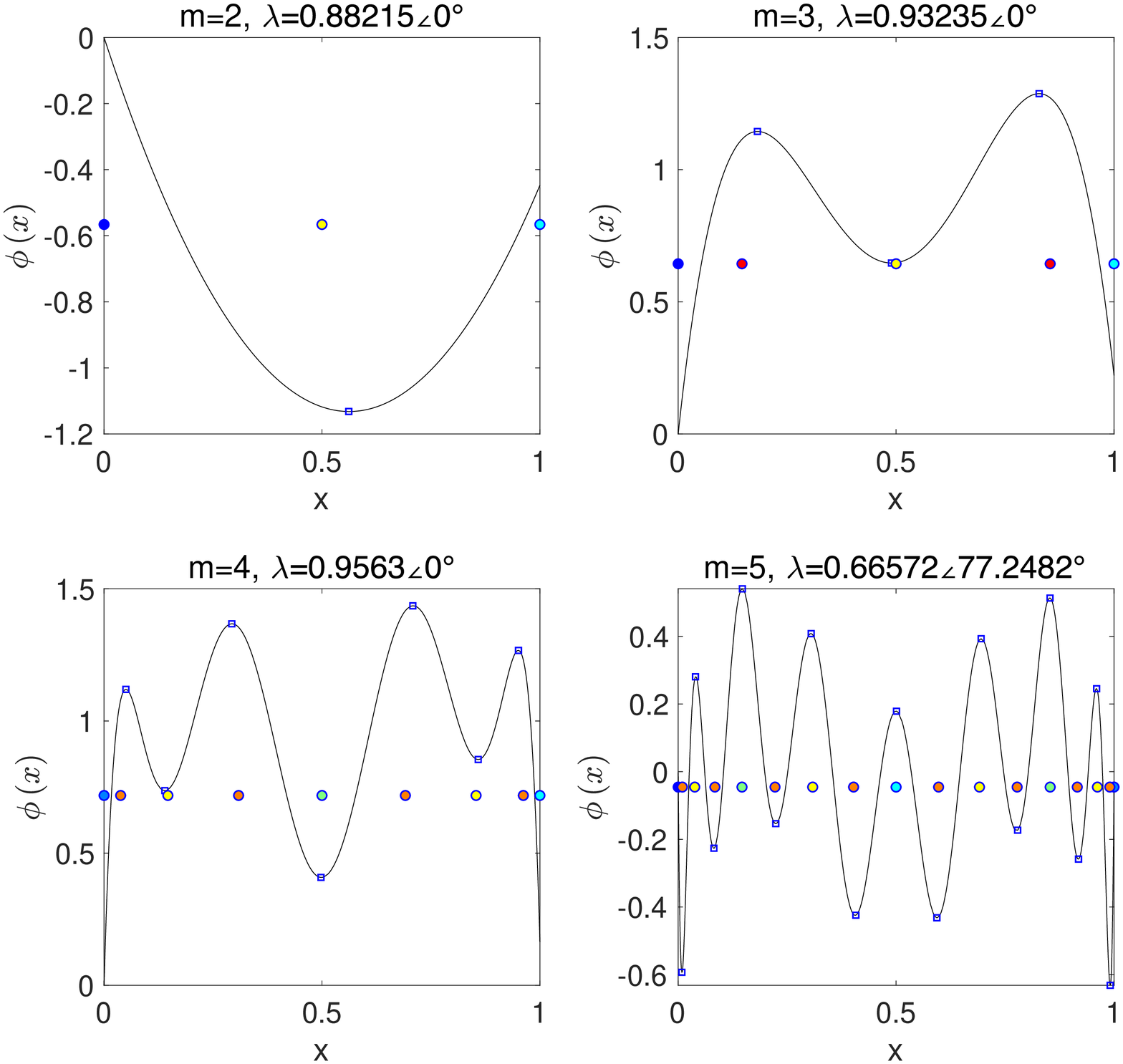}}
		\centerline{(b) logistic map \equref{eq:logistic}}
	\end{minipage}
	\caption{
		Comparison of extremum points (empty squares) and boundary points (colored points, indicating different levels): eigenvalues $\lambda$ (with moduli closest to 1, expressed in modulus and angle) and eigenfunctions (black lines) of the Koopman operator ($n=1000, m=2,3,4,5$) computed with the natural basis for (a) the tent map \equref{eq:tent} and (b) the logistic map \equref{eq:logistic}.\label{fig:Koopman_eigen_natural_boundary_n1000}}
\end{figure*}

Similarly, we can use \equref{eq:koop_U2} to calculate the eigenvalues and eigenfunctions of the Koopman operator by constructing natural basis functions. We take $n=1000$ and $m=2,3,4,5$, and plot the most relevant eigenfunctions, as shown in \figref{fig:Koopman_eigen_natural_boundary_n1000}.

It can be found in \figref{fig:Koopman_eigen_natural_boundary_n1000} that the eigenfunction profile in the natural basis becomes sharp and that for each increase in the number of basis functions, the number of extremum points approximately doubles. We consider the relationship between the mapping $T$ and its iterations $T^N$: every time the number $N$ increases by one, the function graphs of $T^N$ doubles the number of undulations by the process of stretching and folding. The eigenfunctions appear to capture this by wiggling more when the number of basis increases.

So what do these eigenvalues and eigenfunctions represent? Before discussing this issue, we take the logistic map as an example to analyze the dynamical process and the "boundary points" in symbolic partition.

\begin{figure}
	\begin{minipage}{0.48\linewidth}
		\centerline{\includegraphics[width=8.5cm]{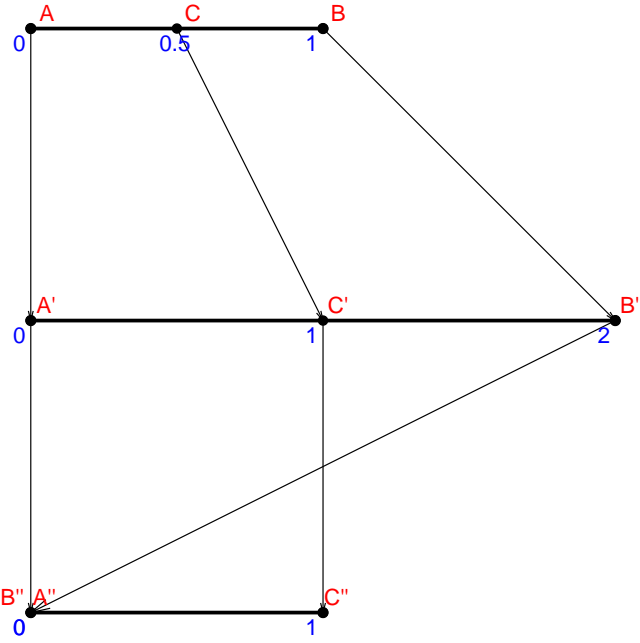}}
	\end{minipage}
	\caption{The dynamical process of the logistic map \equref{eq:logistic} is decomposed into two mappings.\label{fig:logistic_dynamic}}
\end{figure}

The action of the logistic map can be viewed as a process of stretching and refolding on the interval [0,1], as shown in \figref{fig:logistic_dynamic}, which consists of two steps: the first is to stretch the $AB$ interval to $A'B'$ with the points $0$,$0.5$,$1$ mapped to $0$,$1$,$2$. The second folds the $A'B'$ interval to $A''C''$, and maps the points $0$,$1$,$2$ back to $0$,$1$,$0$ respectively. In this process, there are some key "turning points", such as $x=0.5$ of the first mapping and $x=1$ of the second mapping, and the properties of the function are symmetrical with repect to these points.

\begin{figure}
	\begin{minipage}{0.48\linewidth}
		\centerline{\includegraphics[width=8.5cm]{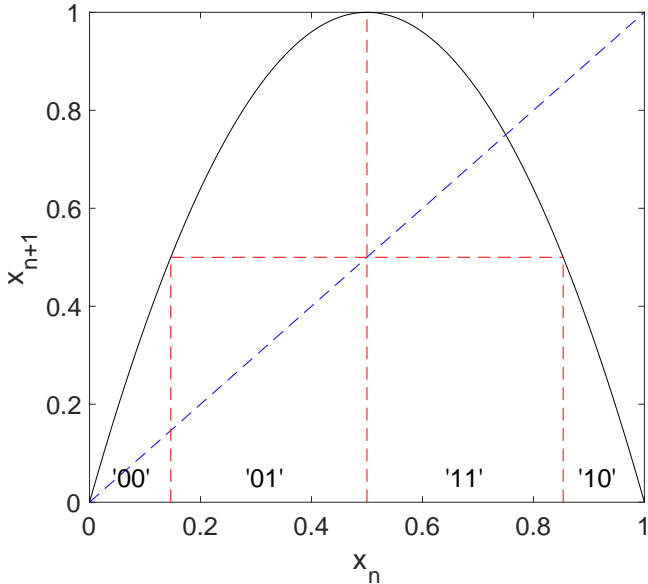}}
	\end{minipage}
	\caption{Symbolic partition of the logistic map \equref{eq:logistic}.\label{fig:logistic_symbolic}}
\end{figure}

We can use symbolic dynamics to describe the folding process of the logistic map, as shown in \figref{fig:logistic_symbolic}. We regard the $x=\frac{1}{2}$ as a "critical point", and denote the areas as "0" and "1" respectively to the left or the right of the critical point. For any point $x\in [0,\frac{1}{2}]$ in the "0" region, its image under the map is a point that lies either in region "0" or in region "1", according to which we can further divide the "0" region into "00" and "01" subregions. Similarly, region "1" can be divided into "11" and "10" subregions. If multiple iterations are considered, we can mark the phase space area with arbitrary precision all the way down to individual point in the limit of infinite sequence of symbols.

In the previous discussion, a key problem is how to find "boundary points". In the logistic map, $x=\frac{1}{2}$ is a "boundary point". From its preimages, we locate two boundary points for the next level $x=\frac{1}{2}\pm\frac{\sqrt{2}}{4}$, which further divides the "0" and "1" region into subregions. From the portraits of the eigenfunction in \figref{fig:Koopman_eigen_Gauss_boundary_n1000m50} and \figref{fig:Koopman_eigen_natural_boundary_n1000}, we see that the extremum points in the profiles are lying close to the "boundary points" of the symbolic partition. This is the first evidence that properly selected eigenfunctions of the Koopman operator may mirror these boundary points.

The boundary points of the tent map and logistic map can be obtained by reverse iteration of the dynamical system, and for both maps there are two preimages for each point. We calculate the boundary points level by level, and list them according to their levels in \figref{fig:Koopman_eigen_Gauss_boundary_n1000m50}. The properly selected eigenfunctions of the Koopman operator are plotted with the Gaussian basis in \figref{fig:Koopman_eigen_Gauss_boundary_n1000m50} and with the natural basis in \figref{fig:Koopman_eigen_natural_boundary_n1000}. To guide the vision, we mark the extrema of the eigenfunctions and the boundary points in these plots, and further list their coordinates in \tabref{tab:map_boundary}.

\begin{table}
	\caption{Comparison of extremum points ($m=2,3,4,5$) and boundary points for different level $l$ as plotted in \figref{fig:Koopman_eigen_natural_boundary_n1000}.\label{tab:map_boundary}}
	\begin{ruledtabular}
		\begin{tabular}{cccccc}
			\multirow{2}{*}{$m$}
			&\multirow{2}{*}{$l$}
			&\multicolumn{2}{c}{tent map \equref{eq:tent}}
			&\multicolumn{2}{c}{logistic map \equref{eq:logistic}}\\ \cline{3-6}
			&&\textrm{extremum}&\textrm{boundary}&\textrm{extremum}&\textrm{boundary}\\
			\colrule
			\multirow{1}{*}{$2$}& \multirow{1}{*}{$0$} & 0.5000 & 0.5000 & 0.5585 & 0.5000\\
			\colrule
			\multirow{3}{*}{$3$}& \multirow{1}{*}{$0$} & 0.5000 & 0.5000 & 0.4868 & 0.5000\\ \cline{2-6}
			& \multirow{2}{*}{$1$} & 0.2502 & 0.2500 & 0.1847 & 0.1464\\
			& & 0.7500 & 0.7500 & 0.8299 & 0.8536\\
			\colrule
			\multirow{7}{*}{$4$}& \multirow{1}{*}{$0$} & 0.5000 & 0.5000 & 0.4972 & 0.5000\\ \cline{2-6}
			& \multirow{2}{*}{$1$} & 0.2503 & 0.2500 & 0.1378 & 0.1464\\
			& & 0.7500 & 0.7500 & 0.8602 & 0.8536\\ \cline{2-6}
			& \multirow{4}{*}{$2$} & 0.1305 & 0.1250 & 0.0518 & 0.0381\\
			& & 0.3750 & 0.3750 & 0.2938 & 0.3087\\
			& & 0.6191 & 0.6250 & 0.7107 & 0.6913\\
			& & 0.8749 & 0.8750 & 0.9485 & 0.9619\\
			\colrule
			\multirow{15}{*}{$5$}& \multirow{1}{*}{$0$} & 0.5000 & 0.5000 & 0.4995 & 0.5000\\ \cline{2-6}
			& \multirow{2}{*}{$1$} & 0.2500 & 0.2500 & 0.1472 & 0.1464\\
			& & 0.7443 & 0.7500 & 0.8522& 0.8536\\ \cline{2-6}
			& \multirow{4}{*}{$2$} & 0.1195 & 0.1250 & 0.0402 & 0.0381\\
			& & 0.3787 & 0.3750 & 0.3024 & 0.3087\\
			& & 0.6278 & 0.6250 & 0.6943 & 0.6913\\
			& & 0.8750 & 0.8750 & 0.9594 & 0.9619\\ \cline{2-6}
			& \multirow{8}{*}{$3$} & 0.0597 & 0.0625 & 0.0087 & 0.0096\\
			& & 0.1894 & 0.1875 & 0.0818 & 0.0842\\
			& & 0.3033 & 0.3125 & 0.2226 & 0.2222\\
			& & 0.4268 & 0.4375 & 0.4055 &0.4024\\
			& & 0.5625 & 0.5625 & 0.5942 & 0.5975\\
			& & 0.6861 & 0.6875 & 0.7764 & 0.7778\\
			& & 0.8172 & 0.8125 & 0.9182 & 0.9157\\
			& & 0.9560 & 0.9375 & 0.9912 & 0.9904\\
			\colrule
		\end{tabular}
	\end{ruledtabular}
\end{table}

For different numbers of Gaussian basis functions as shown in \figref{fig:Koopman_eigen_Gauss_boundary_n1000m50}, the local extremum points of eigenfunctions increases with the number of basis functions. When the number of basis functions is small, the approximation is rough, so the extremum points of eigenfunctions are not in good agreement with boundary points. However, when we increase the number of basis functions, we find that the extremum points of eigenfunctions become increasingly consistent with the boundary points. This is because the added basis functions provide more dynamical undulation details to the eigenfunctions, which then more accurately depict the boundary points.

For different numbers of natural basis functions as shown in \figref{fig:Koopman_eigen_natural_boundary_n1000}, we see that the number of extremum points of eigenfunctions is doubled when the number of basis functions increases by 1. By comparing the coordinates in \tabref{tab:map_boundary}, we find that the newly emerged extremum points correspond to the boundary points at the next level, and with the increase of the basis, the discrepancy between extremum and boundary points also decreases.

We have thus preliminarily verified the consistency between the extremum points of eigenfunctions and the boundary points of the symbolic partition. In the previous discussion, we learned that these boundary points are actually the critical points of symbolic dynamics. We believe that this rule applies to other cases as long as the chaotic dynamics is originated from local stretching and global folding. As we increase the number of basis functions, more partition points at different levels may be identified with greater accuracy, thus enabling us to have a better understanding of the characteristics of the dynamics.

\subsubsection{Robustness of the partition in the presence of noise}
\begin{figure*}
	\begin{minipage}{0.48\linewidth}
		\centerline{\includegraphics[width=8.5cm]{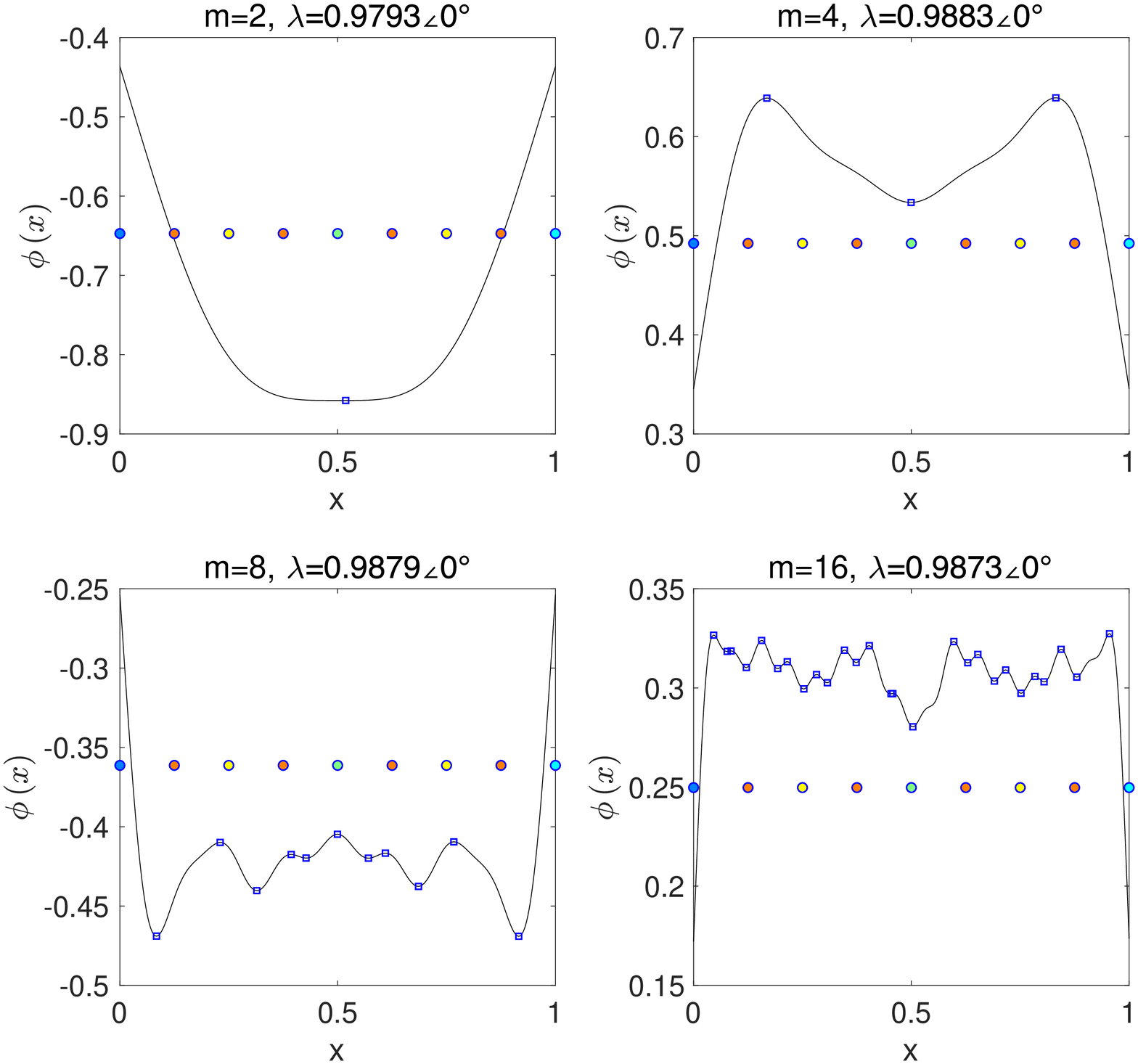}}
		\centerline{(a) tent map \equref{eq:tent}}
	\end{minipage}
	\hfill
	\begin{minipage}{0.48\linewidth}
		\centerline{\includegraphics[width=8.5cm]{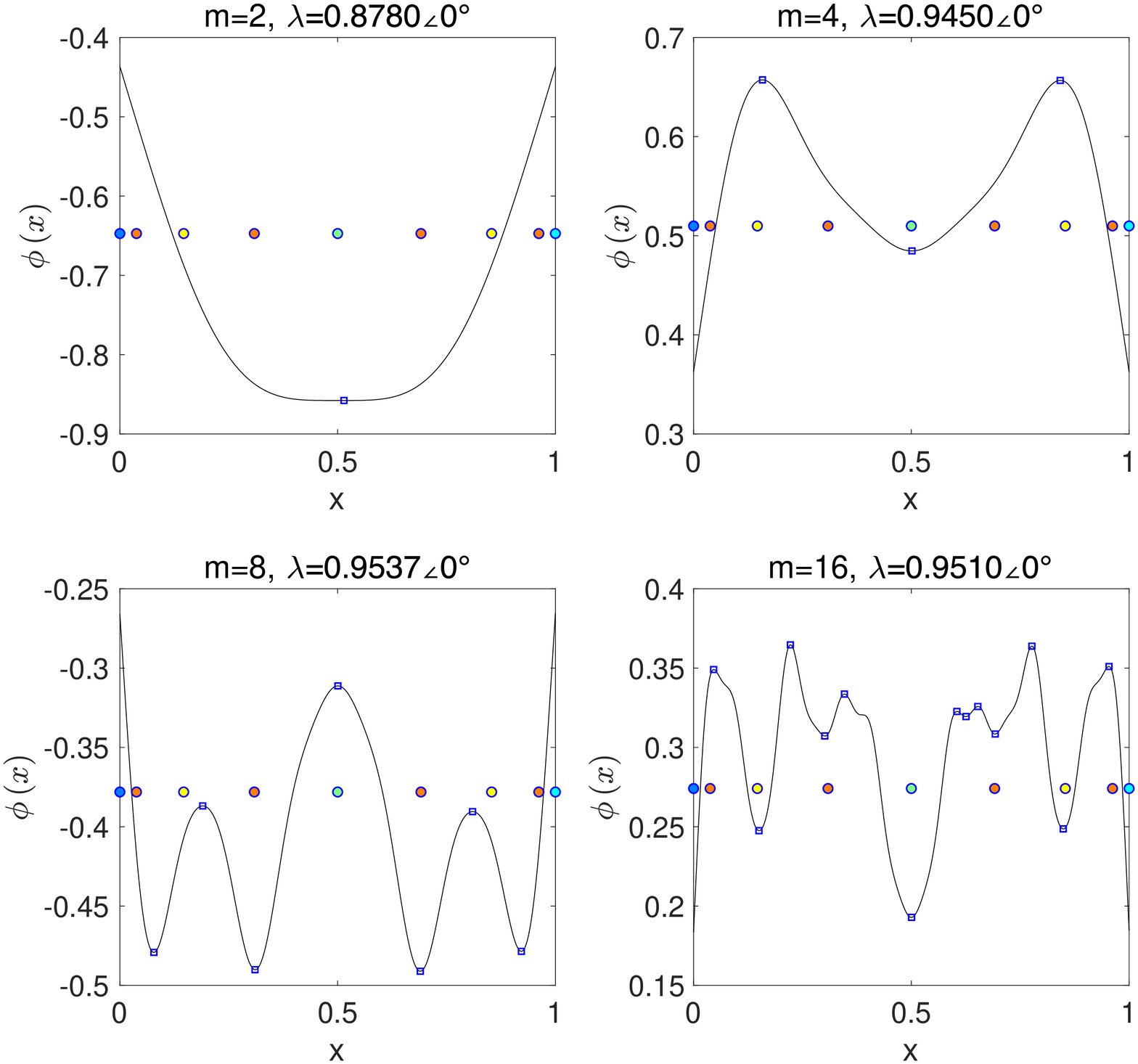}}
		\centerline{(b) logistic map \equref{eq:logistic}}
	\end{minipage}
	\caption{Same as in \figref{fig:Koopman_eigen_Gauss_boundary_n1000m50}, but subject to Gaussian white noise with variance $\sigma=0.001$, averaged over $50$ realizations.\label{fig:Koopman_eigen_noise_n1000}}
\end{figure*}

In the real world, the evolution of a nonlinear system is often contaminated by the omnipresent noise. Heninger proposed the finest state-space resolution that can be achieved in a physical dynamical system is limited by noise \cite{heninger2015neighborhoods}, which in fact revives a form of perturbation theory in \cite{heninger2018perturbation}. The finite approximation of the Koopman operator presented above of course brings error in the computation but at the same time grants robustness even in the presence of noise.  We will investigate what kind of role noise plays and how much the spectral properties of the Koopman operator are altered. For simplicity, the noise is assumed to be Gaussian white. The parameters in the following computation are: the number of points $n=1000$, the number of Gaussian functions $m=2,4,8,16$, the noise variance $\sigma=0.001$, and also an average of $50$ realizations of $L$ from the same $K$ is used for the construction of \equref{eq:koop_U2}. Similar procedure is used for the computation as before, and we take the eigenvalues closest to 1 to plot the corresponding eigenfunctions in \figref{fig:Koopman_eigen_noise_n1000}. 

Comparing \figref{fig:Koopman_eigen_Gauss_boundary_n1000m50} with \figref{fig:Koopman_eigen_noise_n1000}, we find that the set of boundary points by Koopman operator is almost the same, which shows the robustness of our scheme.

\subsubsection{Consistency in different truncations}
With the increase of the number of basis functions, the eigenfunctions describe the boundary points more finely: the more the basis functions, the higher the level and fineness of the boundary points prescribed by the eigenfunctions. In order to verify our conclusion, we make the following comparison: when the number of basis functions doubles, we explore the relationship between the extremum points of pertinent eigenfunctions. In order to compare the similarity in dynamical characteristics, we define a correlation coefficient of the eigenfunctions
\begin{equation}
\rho(\phi_1,\phi_2)=\dfrac{\sum_i(\phi_{1i}-\overline{\phi}_1)(\phi_{2i}-\overline{\phi}_2)}{\sqrt{(\sum_i(\phi_{1i}-\overline{\phi}_1)^2)(\sum_i(\phi_{2i}-\overline{\phi}_2)^2)}}
\,,
\label{eq:corr}
\end{equation}
where $\phi_1,\phi_2$ are the corresponding eigenfunctions, and $\overline{\phi}_1,\overline{\phi}_2$ are their mean values.

To ensure the consistency of corresponding eigenfunctions for different truncations, the following scheme is designed to search $\phi_{m_2}$ when $\phi_{m_1}$ is known
\begin{equation}
\mathop{\arg\min}_{\boldsymbol{\phi_{m_2}}} \ \ || U\phi_{m_2}-\lambda \phi_{m_2} || + \mu |\rho(\phi_{m_1},\phi_{m_2})|
\,,
\label{eq:argmin}
\end{equation}
where $U$ is the Koopman operator and $\lambda$ is the eigenvalue of $\phi_{m_1}$. The first term in \equref{eq:argmin} is the definition of eigenfunction and the second term is the correlation coefficient defined in \equref{eq:corr}. $\mu$ is a relative weight. Generally speaking, we should ensure first the correctness of the eigenfunction, so usually $\mu\ll 1$.

\begin{figure*}
	\begin{minipage}{0.48\linewidth}
		\centerline{\includegraphics[width=8.5cm]{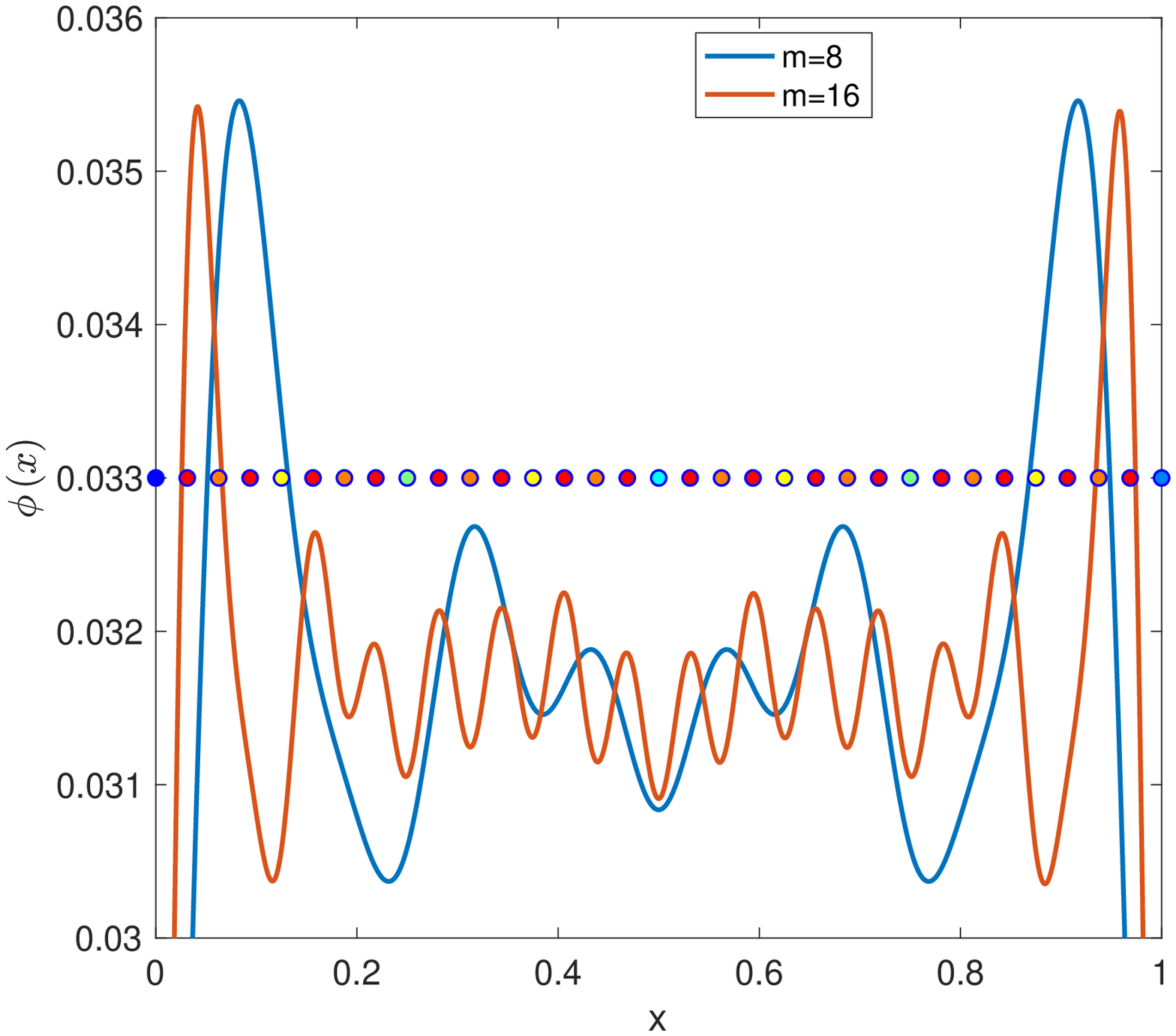}}
		\centerline{(a) tent map \equref{eq:tent}}
	\end{minipage}
	\hfill
	\begin{minipage}{0.48\linewidth}
		\centerline{\includegraphics[width=8.5cm]{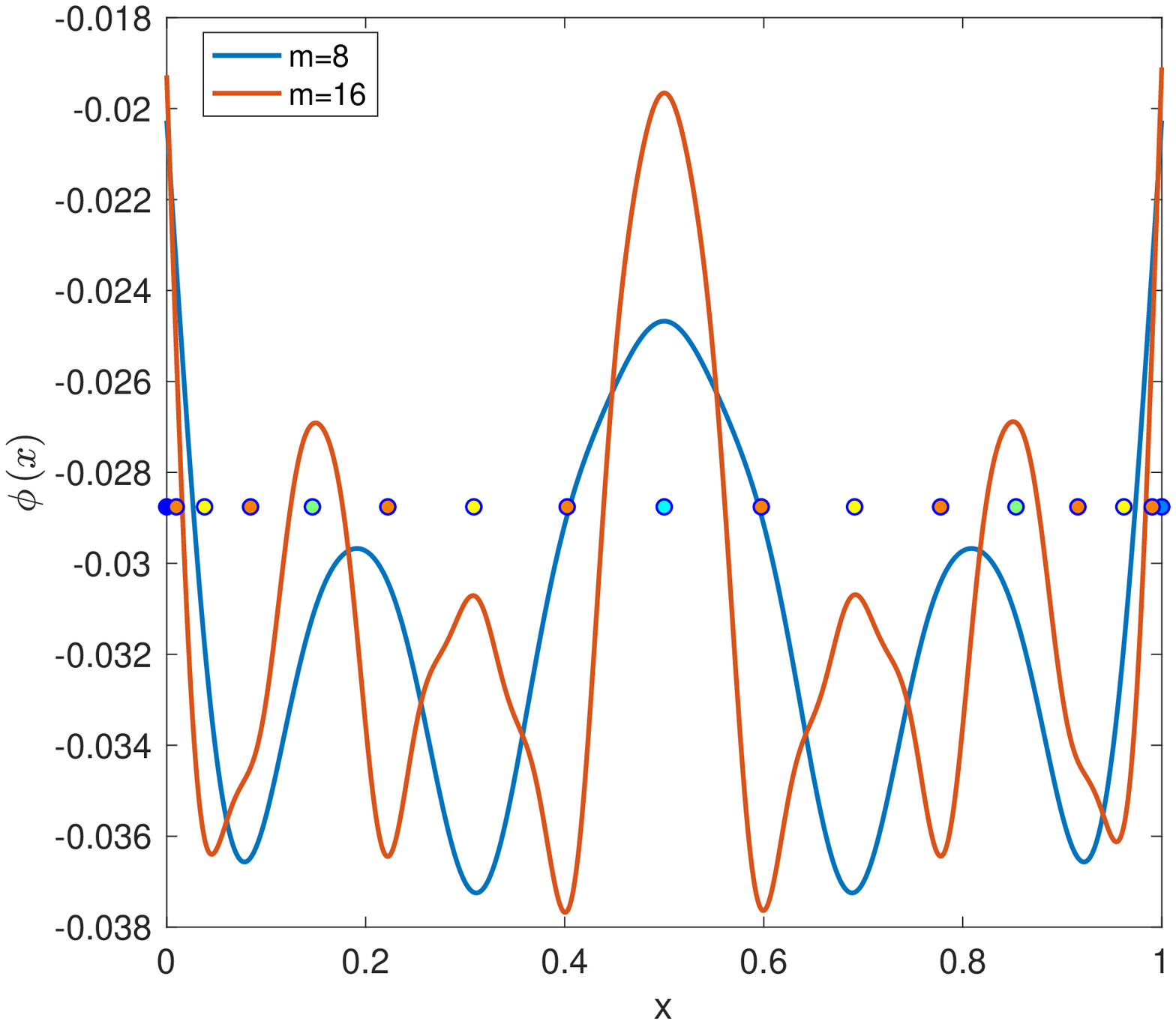}}
		\centerline{(b) logistic map \equref{eq:logistic}}
	\end{minipage}

	\caption{Correspondence between eigenfunctions with different $m$ ($n=1000$): the resulted eigenfunction with $m=16$ (red) starting from its counterpart with $m=8$ (blue) based on \equref{eq:argmin} ($\mu=0.01$), for (a) the tent map \equref{eq:tent} and (b) the logistic map \equref{eq:logistic}. \label{fig:Koopman_findeigen_m8m16}}
\end{figure*}

We did the minimization for $m_1=8$ and $m_2=16$ and in \figref{fig:Koopman_findeigen_m8m16} compared the results (blue curves for $m_1=8$ and red curves for $m_2=16$).  As the number of basis functions doubles, the number of extremum points also doubles, but the latter eigenfunction keeps all the extremum points of the previous one. Through perusing and comparing data, we find that the extra extremum points are the boundary points of the system at higher levels, which shows that as the number of basis functions increases, it is possible to reach higher and higher levels. As long as we can get enough basis functions, we are able to determine symbolic partitions accurate enough.

\subsubsection{One-dimensional maps with multiple peaks}

We have successfully partitioned the phase space of 1-d unimodal maps with eigenfunctions of the Koopman operator, but its applicability to other systems needs further investigation. Here we discuss certain multimodal maps, as given in \equref{eq:tents_f1} and \equref{eq:tents_f2}.

\begin{subequations}
	\begin{align}
	f_1(x)&=\begin{cases}
	4x,&(0\leqslant x\leqslant 0.25)\\
	2-4x,&(0.25\leqslant x\leqslant 0.5)\\
	4x-2,&(0.5\leqslant x\leqslant 0.75)\\
	3-4x,&(0.75\leqslant x\leqslant 1)
	\end{cases}\label{eq:tents_f1}\\
	f_2(x)&=\begin{cases}
	4x,&(0\leqslant x\leqslant 0.25)\\
	2-4x,&(0.25\leqslant x\leqslant 0.5)\\
	2x-1,&(0.5\leqslant x\leqslant 0.75)\\
	1.5-2x,&(0.75\leqslant x\leqslant 1)\label{eq:tents_f2}
	\end{cases}
	\end{align}
\label{eq:tents}%
\end{subequations}

\begin{figure*}
	\begin{minipage}{0.48\linewidth}
		\centerline{\includegraphics[width=8.5cm]{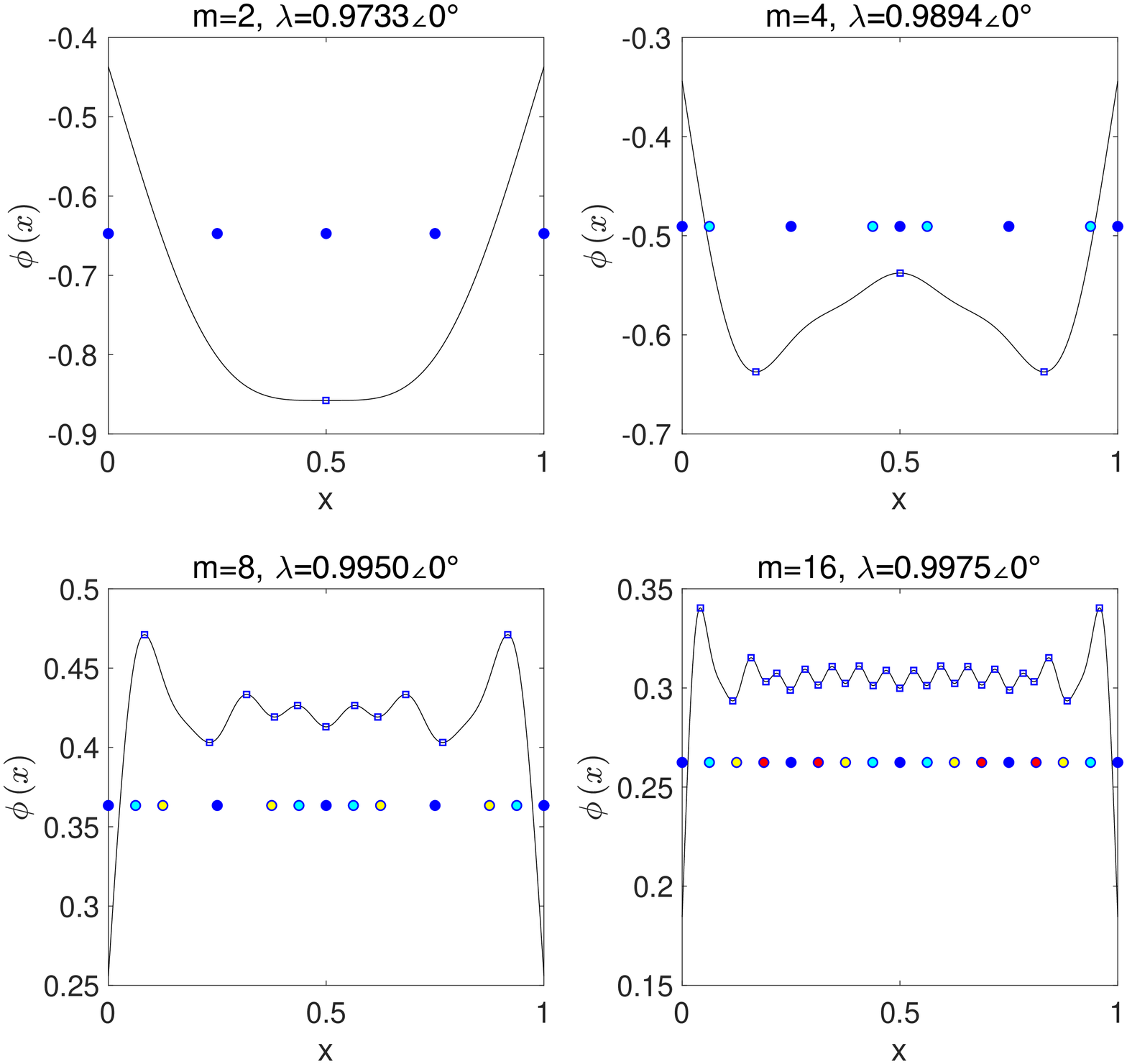}}
		\centerline{(a) $f_1$ map \equref{eq:tents_f1}}
	\end{minipage}
	\hfill
	\begin{minipage}{0.48\linewidth}
		\centerline{\includegraphics[width=8.5cm]{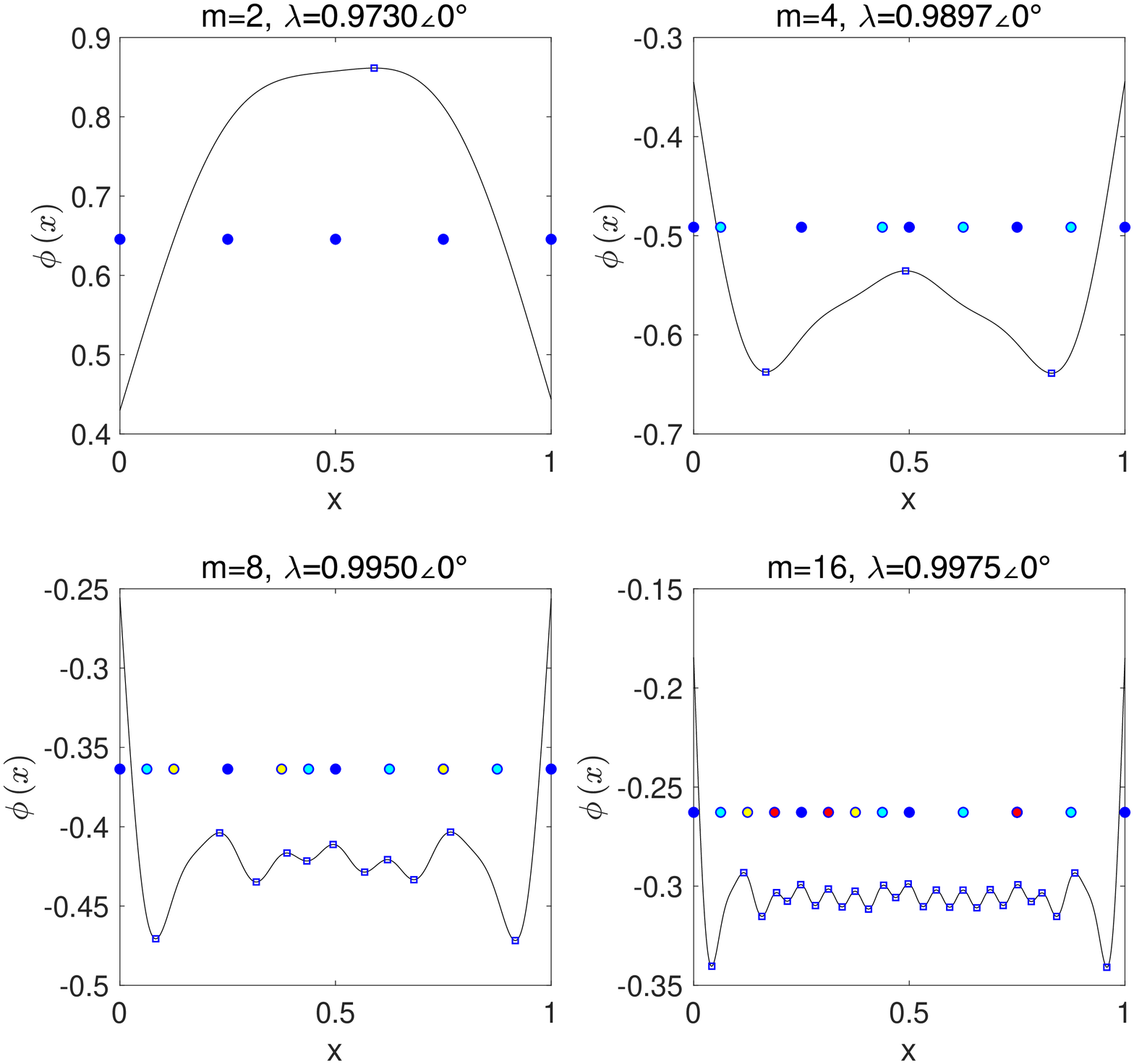}}
		\centerline{(b) $f_2$ map \equref{eq:tents_f2}}
	\end{minipage}
	\caption{
		Comparison of extremum points (empty squares) and boundary points (colored points, indicating different levels): eigenvalues $\lambda$ (for the first eigenvalue, expressed in modulus and angle) and eigenfunctions (black lines) of the Koopman operator ($n=1000, m=2,4,8,16$) computed with the Gaussian basis for (a) $f_1$ map \equref{eq:tents_f1} and (b) $f_2$ map \equref{eq:tents_f2}.
		\label{fig:Tents_eigen_n1000m50}}
\end{figure*}

\figref{fig:Tents_eigen_n1000m50} displays the relevant eigenfunctions of the Koopman operator for $f_1$ and $f_2$. In both maps, there are three critical points: $0.25$, $0.5$ and $0.75$ at the first level. The positions of the boundary points seem to coincide quite well with those of the extremum points of the corresponding eigenfunction.

\begin{figure*}
	\begin{minipage}{0.48\linewidth}
		\centerline{\includegraphics[width=8.5cm]{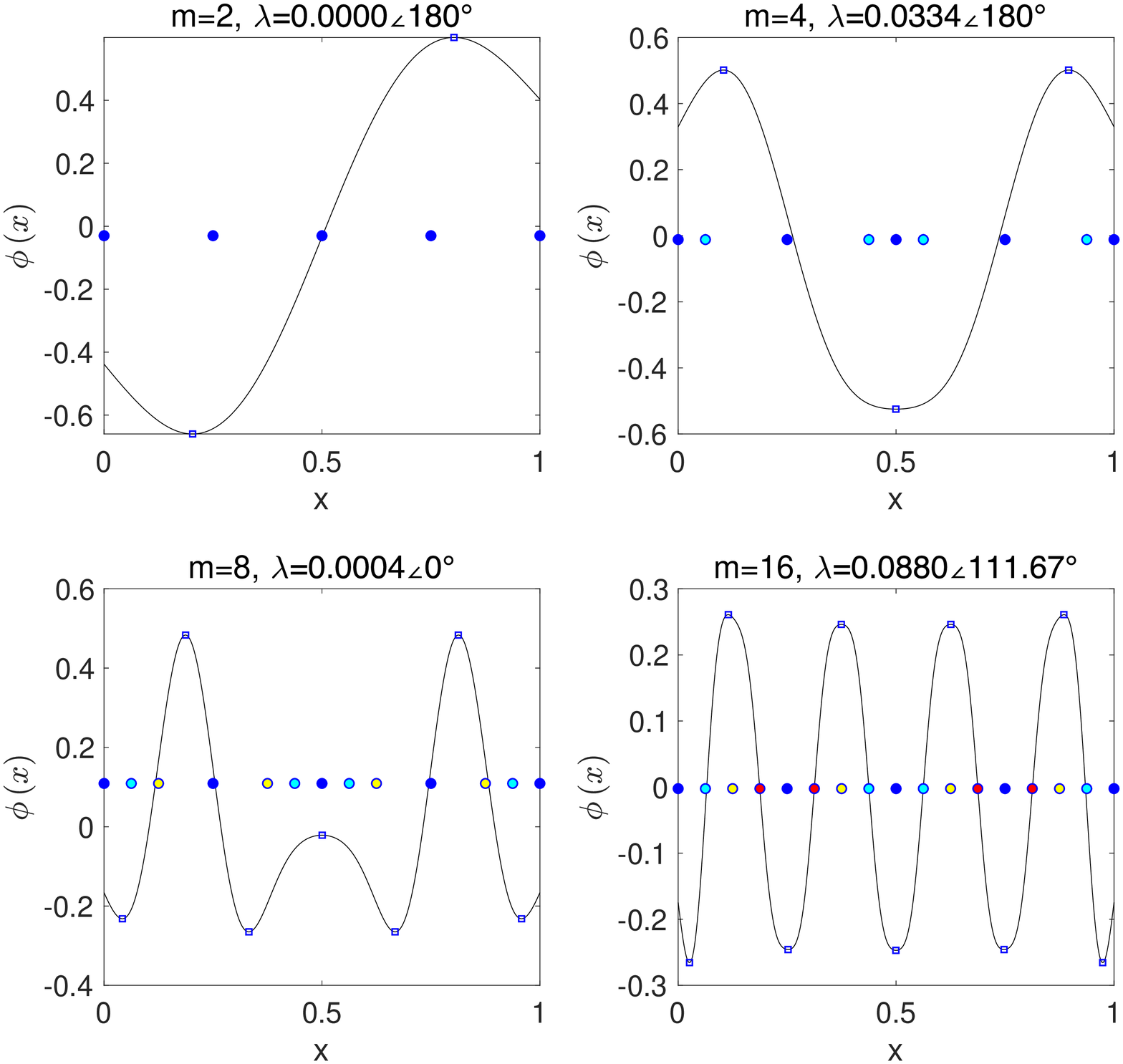}}
		\centerline{(a) $f_1$ map \equref{eq:tents_f1}}
	\end{minipage}
	\hfill
	\begin{minipage}{0.48\linewidth}
		\centerline{\includegraphics[width=8.5cm]{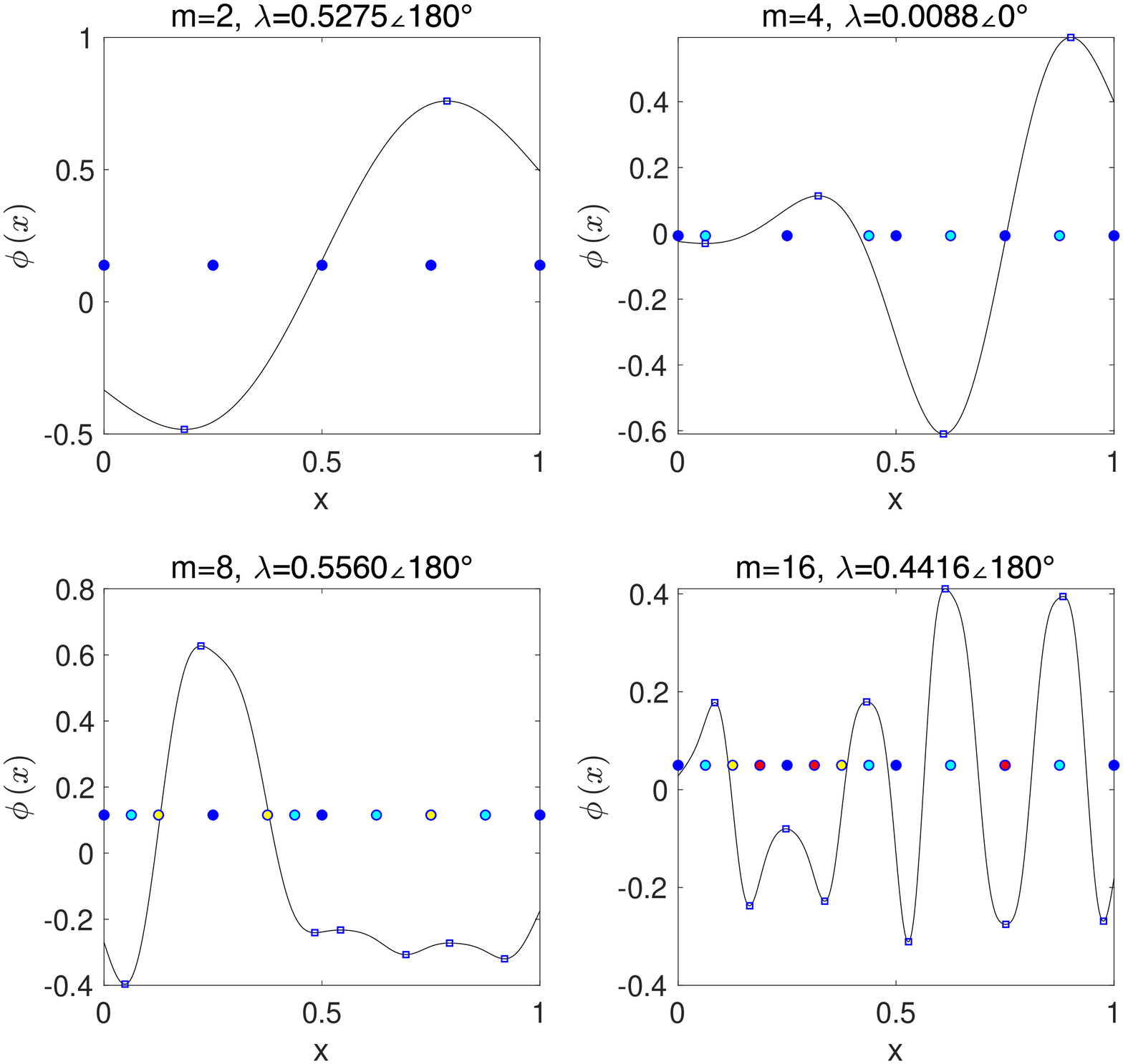}}
		\centerline{(b) $f_2$ map \equref{eq:tents_f2}}
	\end{minipage}
	\caption{
		Comparison of extremum points (empty squares) and boundary points (colored points, indicating different levels): eigenvalues $\lambda$ (for the second eigenvalue, expressed in modulus and angle) and eigenfunctions (black lines) of the Koopman operator ($n=1000, m=2,4,8,16$) computed with the Gaussian basis for (a) $f_1$ map \equref{eq:tents_f1} and (b) $f_2$ map \equref{eq:tents_f2}.
		\label{fig:Tents_eigen_n1000m50_choose2}}
\end{figure*}

However, there are some differences in the two maps. For example, in map $f_2$, the function values at $x=0.25$ and $x=0.75$ are twice as different, but the eigenfunctions of the two maps are surprisingly similar, which had made us doubt the accuracy of our computation. By checking more eigenfunction plots, we find that this is true only for the principle eigenfunction (with eigenvalue closest to 1). If we plot the second most relevant function, as in \figref{fig:Tents_eigen_n1000m50_choose2}, the two eigenfunction profiles are very different. This shows that other eigenfunctions can describe dynamical difference between the two, and the eigenfunctions for disparate eigenvalues contain different information of the dynamics.

\subsection{Two-dimensional map: H\'{e}non map}

H\'{e}non map is a discrete mapping system generating chaos with appropriate parameters \cite{simo1979henon}, which reads
\begin{equation}
\begin{aligned}
x_{n+1}&=y_n+1-ax_n^2\\
y_{n+1}&=bx_n
\end{aligned}\label{eq:Henon}
\end{equation}
and has two fixed points: $x^*=\dfrac{b-1\pm\sqrt{(b-1)^2+4a}}{2a},y^*=bx^*$. For the parameter value $a=1.4$ and $b=0.3$, H\'{e}non map produces chaos and a strange attractor appears in the phase space.

\begin{figure*}
	\begin{minipage}{0.96\linewidth}
		\centerline{\includegraphics[width=17cm]{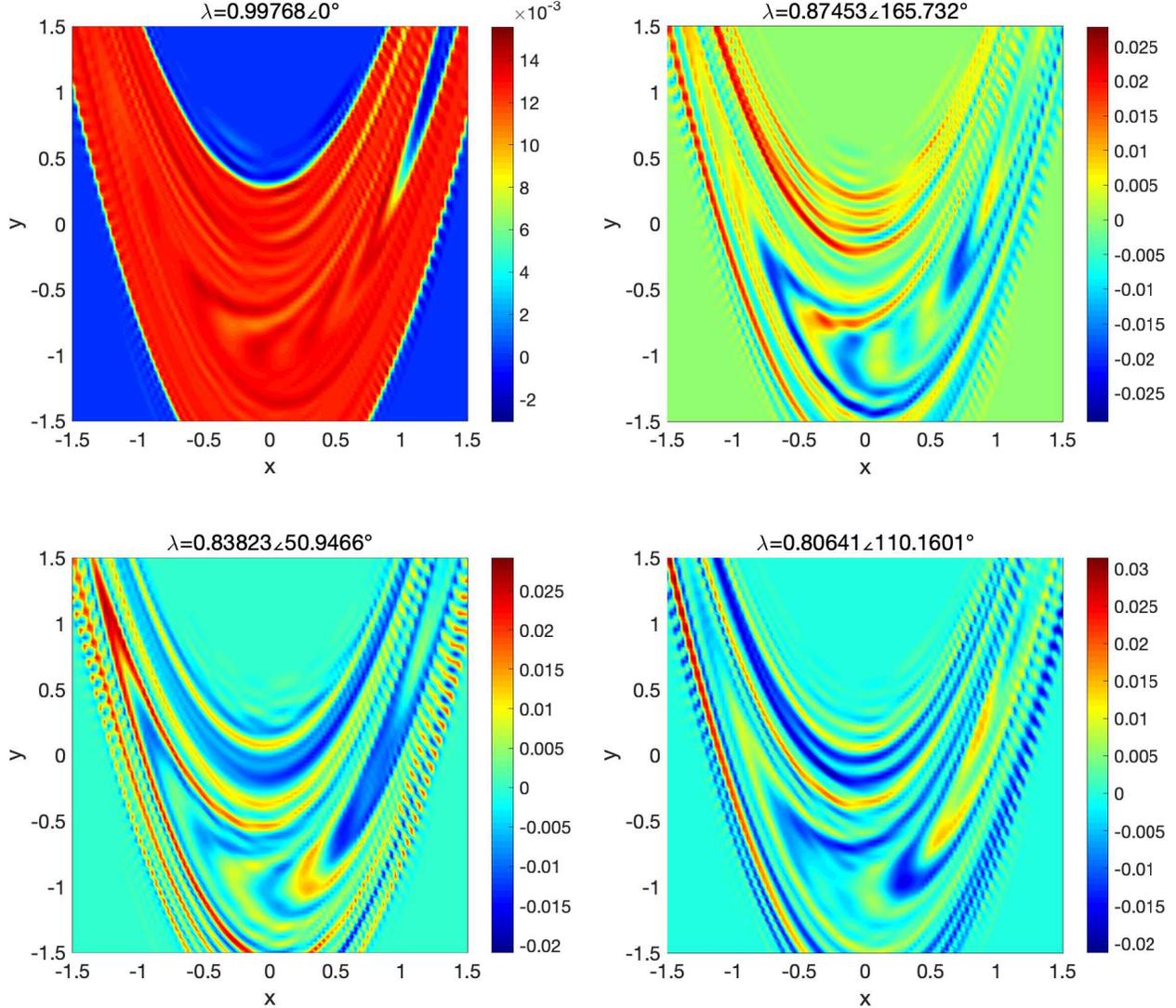}}
	\end{minipage}
	\caption{Eigenvalues $\lambda$ (the first four eigenvalues with moduli closest to 1, expressed in modulus and angle) and eigenfunctions (color plots) of H\'{e}non map \equref{eq:Henon} ($a=1.4, b=0.3$) in the rectangular area $[1.5\times 1.5]$ with the Gaussian basis ($n=100^2$,$m=50^2$, $d_j=3/45$).\label{fig:Henon_eigen}}
\end{figure*}

In a two-dimensional map, the calculation of the Koopman operator eigenfunction is slightly different from the one-dimensional counterpart. For example, we may use two-dimensional Gaussian functions as basis
\begin{equation}
g(x,y)=Cexp\left({\dfrac{-(x-x_j)^2-(y-y_j)^2}{2d_j^2}}\right)
\,,
\end{equation}
where $C$ is a normalization constant, $(x_j,y_j)$ is the center of the Gaussian wave packet, and $d_j$ is the width of the wave packet. In the rectangular area $[-1.5,1.5]\times [-1.5,1.5]$ which contains the attractor, we deploy the Gaussian basis functions with centers uniformly distributed on a $50\times 50$ lattice and a wave packet width $d_j=3/45$, and with the number of points $n=100\times 100$ calculate the eigenvalues and eigenfunctions. We take the first four eigenfunctions whose eigenvalues are closest to 1. When the eigenfunction value is complex, the real part is used for the plot. \figref{fig:Henon_eigen} displays the color images of the eigenfunctions. Note that the profile of the eigenfunction is very similar to that of the stable manifold of H\'{e}non map \cite{grassberger1985generating}. The boundary points of the H{\' e}non map could be chosen from the homoclinic tangencies of the stable and unstable manifold~\cite{jaeger1997structure,grassberger1985generating}. Here, the asymptotic state of a typical point is on the closure of the unstable manifold while the Koopman eigenfunctions are capable of revealing the stable manifold, which partly explains why the partition boundaries could be located in this new approach.

Alternatively the Koopman operator may be approximated with the natural function basis. However, it is slightly different from the one-dimensional case and involves a minor complication. In the two-dimensional map, there are two state observables $x$ and $y$, either of which can be used for the construction. For finite resolution, different observables may emphasize different aspects of the dynamics and thus have different projections on the eigenfunctions. In the discussion of H\'{e}non map, we take the $x-$direction as an example to emphasize the stretching and folding dynamics. At the same time, for this set of parameter values, the H\'{e}non map is dissipative and the strange attractor has a fractal dimension slightly greater than 1. The natural basis is just defined on the attractor and we only plot the eigenfunctions supported on the attractor.

\begin{figure*}
	\begin{minipage}{0.96\linewidth}
		\centerline{\includegraphics[width=17cm]{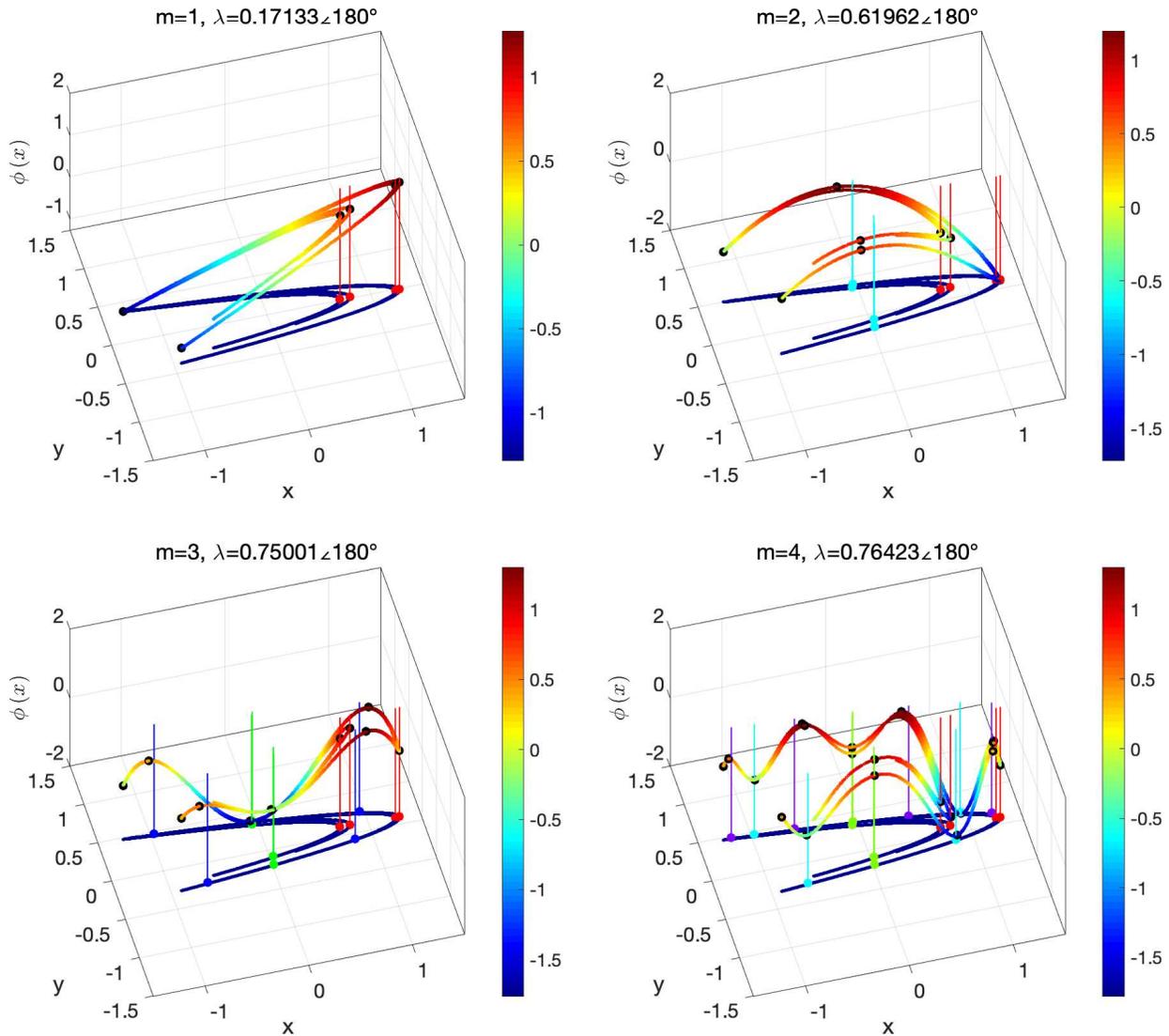}}
	\end{minipage}
	\caption{
		Comparison of extremum points (black dots) and boundary points (colored dots, indicating different levels): eigenvalues $\lambda$ (with moduli closest to 1, expressed in modulus and angle) and eigenfunctions (colored lines) of the Koopman operator ($n=10000,m=1,2,3,4$) computed with the natural basis for the H\'{e}non map \equref{eq:Henon} ($a=1.4, b=0.3$).
		\label{fig:Henon_eigen_natural_attr_n10000}}
\end{figure*}

\figref{fig:Henon_eigen_natural_attr_n10000} shows the eigenfunction plots on the attractor with $n=10000$ and $m=1,2,3,4$. Since the attractor is thin and difficult to represent, we use a three-dimensional plot, portraying the eigenfunction of the Koopman operator with height and color at the same time, and also project it to the $x-y$ plane.

As can be seen from \figref{fig:Henon_eigen_natural_attr_n10000}, with the increase of the number of basis functions, the structure of eigenfunction images becomes increasingly complicated, and the number of extremum points increases exponentially, which is consistent with our conclusion in one-dimensional mapping. For simplicity in a coarse-grained picture, we regard the attractor of H\'{e}non map as consisting of four main segments over which eigenfunctions are plotted. When $m=1$, the eigenfunction identifies only the principle group of critical points, one on each main segment and dividing the attractor of the H\'{e}non map into two parts. When $m=2$, there is an extra set of extremum points, which divide each subsegment from the previous step into two parts, which is very similar to the 1-d case.

\begin{figure}
	\centering
	\includegraphics[scale=0.4]{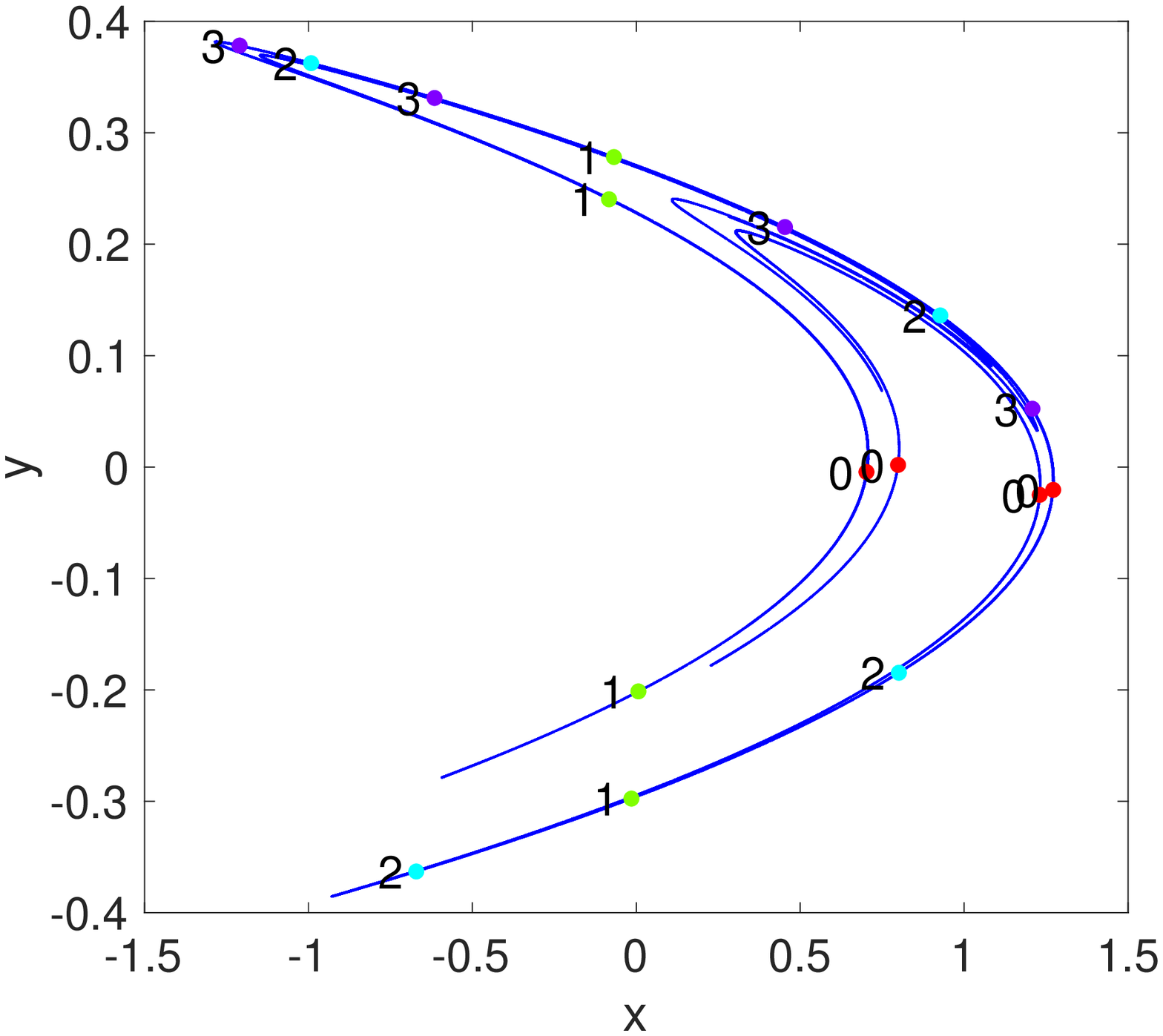}
	\caption{Boundary points of H\'{e}non map \equref{eq:Henon}: the principle ones (marked by 0) and their preimages (marked by positive integers for different levels).\label{fig:Henon_boundary}}
\end{figure}

The numerical values of boundary points are given in \cite{jaeger1997structure}, the principle four of which is listed in \tabref{tab:henon_boundary}. They are all located in the main bending region of the attractor, and close to the extremum points identified in the eigenfunction. In order to find the boundary points at other levels, we carry out backward evolution of the four boundary points to obtain their preimages. As depicted in \figref{fig:Henon_boundary}, we label the preimages with positive numbers, and $0$ represents the original four boundary points.

\begin{table}
	\caption{Comparison of extremum points ($m=1,2,3,4$) and boundary points (given in \cite{jaeger1997structure}) at different level $l$ of the H\'{e}non map ($a=1.4,b=0.3$) as plotted in \figref{fig:Henon_eigen_natural_attr_n10000}.\label{tab:henon_boundary}}
	\begin{ruledtabular}
		\begin{tabular}{cccc}
			\textrm{$m$}&\textrm{$l$}&\textrm{extremum}&\textrm{boundary}\\
			\colrule
			\multirow{4}{*}{$1$}& \multirow{4}{*}{$0$} & \pos{\ 0.7079}{\ 0.0120} & \pos{\ 0.7021}{-0.0044} \\
			& & \pos{\ 0.8019}{\ 0.0191} & \pos{\ 0.7986}{\ 0.0019} \\
			& & \pos{\ 1.2326}{-0.0153} & \pos{\ 1.2307}{-0.0249} \\
			& & \pos{\ 1.2729}{-0.0091} & \pos{\ 1.2717}{-0.0205} \\
			\colrule
			\multirow{8}{*}{$2$}& \multirow{4}{*}{$0$} & \pos{\ 0.7052}{-0.0009} & \pos{\ 0.7021}{-0.0044} \\
			& & \pos{\ 0.7995}{\ 0.0041} & \pos{\ 0.7986}{\ 0.0019} \\
			& & \pos{\ 1.2720}{-0.0192} & \pos{\ 1.2307}{-0.0249} \\
			& & \pos{\ 1.2720}{-0.0192} & \pos{\ 1.2717}{-0.0205} \\ \cline{2-4}
			& \multirow{4}{*}{$1$} & \pos{-0.1418}{-0.3100} & \pos{-0.0148}{-0.2976} \\
			& & \pos{-0.1258}{-0.2206} & \pos{\ 0.0065}{-0.2013} \\
			& & \pos{-0.2164}{\ 0.2936} & \pos{-0.0832}{\ 0.2403} \\
			& & \pos{-0.2164}{\ 0.2936} & \pos{-0.0682}{\ 0.2782} \\
			\colrule
			\multirow{12}{*}{$3$}& \multirow{4}{*}{$0$} & \pos{\ 0.7045}{-0.0026} & \pos{\ 0.7021}{-0.0044} \\
			& & \pos{\ 0.7995}{\ 0.0041} & \pos{\ 0.7986}{\ 0.0019} \\
			& & \pos{\ 1.2713}{-0.0217} & \pos{\ 1.2307}{-0.0249} \\
			& & \pos{\ 1.2713}{-0.0217} & \pos{\ 1.2717}{-0.0205} \\ \cline{2-4}
			& \multirow{4}{*}{$1$} & \pos{-0.0330}{-0.2981} & \pos{-0.0148}{-0.2976} \\
			& & \pos{-0.0121}{-0.2041} & \pos{\ 0.0065}{-0.2013} \\
			& & \pos{-0.0982}{\ 0.2431} & \pos{-0.0832}{\ 0.2403} \\
			& & \pos{-0.0862}{\ 0.2802} & \pos{-0.0682}{\ 0.2782} \\ \cline{2-4}
			& \multirow{4}{*}{$2$} & \pos{-1.0459}{\ 0.3549} & \pos{-0.9918}{\ 0.3625} \\
			& & \pos{-0.7503}{-0.3700} & \pos{-0.6711}{-0.3630} \\
			& & \pos{\ 0.9130}{-0.1574} & \pos{\ 0.8011}{-0.1846} \\
			& & \pos{\ 1.0081}{\ 0.1176} & \pos{\ 0.9275}{\ 0.1361} \\
			\colrule
			\multirow{16}{*}{$4$}& \multirow{4}{*}{$0$} & \pos{\ 0.7045}{-0.0026} & \pos{\ 0.7021}{-0.0044} \\
			& & \pos{\ 0.8002}{\ 0.0059} & \pos{\ 0.7986}{\ 0.0019} \\
			& & \pos{\ 1.2713}{-0.0217} & \pos{\ 1.2307}{-0.0249} \\
			& & \pos{\ 1.2713}{-0.0217} & \pos{\ 1.2717}{-0.0205} \\ \cline{2-4}
			& \multirow{4}{*}{$1$} & \pos{-0.0086}{-0.2954} & \pos{-0.0148}{-0.2976} \\
			& & \pos{\ 0.0138}{-0.2002} & \pos{\ 0.0065}{-0.2013} \\
			& & \pos{-0.0856}{\ 0.2413} & \pos{-0.0832}{\ 0.2403} \\
			& & \pos{-0.0724}{\ 0.2787} & \pos{-0.0682}{\ 0.2782} \\ \cline{2-4}
			& \multirow{4}{*}{$2$} & \pos{-0.9937}{\ 0.3495} & \pos{-0.9918}{\ 0.3625} \\
			& & \pos{-0.6805}{-0.3638} & \pos{-0.6711}{-0.3630} \\
			& & \pos{\ 0.8132}{-0.1780} & \pos{\ 0.8011}{-0.1846} \\
			& & \pos{\ 0.9338}{\ 0.1347} & \pos{\ 0.9275}{\ 0.1361} \\ \cline{2-4}
			& \multirow{4}{*}{$3$} & \pos{\ 1.2205}{\ 0.0336} & \pos{\ 1.2084}{\ 0.0523} \\
			& & \pos{-1.2334}{\ 0.3797} & \pos{-1.2099}{\ 0.3783} \\
			& & \pos{-0.5413}{\ 0.3240} & \pos{-0.6154}{\ 0.3313} \\
			& & \pos{\ 0.3897}{\ 0.2237} & \pos{\ 0.4535}{\ 0.2154} \\
			\colrule
		\end{tabular}
	\end{ruledtabular}
\end{table}

The eigenfunctions together with the boundary points are plotted in \figref{fig:Henon_eigen_natural_attr_n10000}. The coordinates of extremum points of eigenfunctions and boundary points are listed in \tabref{tab:henon_boundary}. By checking their positions in \figref{fig:Henon_eigen_natural_attr_n10000} and comparing their coordinates in \tabref{tab:henon_boundary}, we find that the extremum points of eigenfunctions always correspond to the boundary points of the symbolic partition of the H\'{e}non map, and higher-level boundary points emerge with the increase of the number of basis functions. As such, the extremum points of the eigenfunction match better with the corresponding boundary points, which is consistent with the one-dimensional case.

\section{Conclusion\label{sec:conclusion}}

Koopman operator describes the evolution of observable functions in phase space, which could be used to excavate important patterns of dynamics. In a spectral decomposition of the Koopman operator, numerous eigenvalues and eigenfunctions are identified but it remains a big challenge to identify the most relevant ones and reveal their dynamical significance. In the current paper, we find that extrema of properly selected eigenfunctions overlap with the boundary points of symbolic partitions in the 1-d or 2-d maps with chaotic dynamics. With well-chosen basis functions, the spectral properties of the Koopman operator could easily be computed and the accuracy increases with the truncation order. Compared to other basis, the natural one directly drawn from the evolution data is the most efficient, which should be applicable to a wide-variety of other non-linear systems. 

Symbolic partition is always a mind-boggling problem in the study of nonlinear dynamics, especially when the phase space is high dimensional~\cite{gonchenko2016variety,gonchenko2014simple,gonchenko2005three}. Most previous approaches are concentrated in the construction of stable and unstable manifolds or identification of some topological index, which are hard to invoke when orbit structures are complicated, being usually the case for coupled nonlinear systems. In the current paper, a totally different route is taken by considering the spectral properties of the Koopman operator, thus avoiding the nearly impossible description of possibly intricate geometric features in the phase space. Therefore, it is reasonable and desirable to extend our new scheme to the treatment of complex nonlinear systems and achieve proper symbolic description.

Although the analysis is demonstrated in the well-know 1-d or 2-d maps, further investigation is still needed for its extension, especially in the selection of proper eigenfunctions. In this paper, eigenfunctions with eigenvalues close to $1$ are chosen for the symbolic partition. It would be interesting to check the roles of other eigenfunctions and propose a systematic scheme for the selection of relevant eigenfunctions. We also checked and compared eigenfunctions for different truncations in the computation and found that the extremum points are consistent in these eigenfunctions but new ones emerge with an increasing number of basis functions, unfolding higher levels of partition boundaries. Therefore, it seems possible to utilize the truncation order to coarsen or refine the description when necessary, which should be explored in more detail in future.    

In a 1-d example in the above discussion, small Gaussian white noise is added to the evolution but the partition based on the Koopman operator seems intact, which shows the robustness of the current scheme. It would be interesting to check if and how the partition changes with the noise intensity since there should exist an optimal partition for noise with a finite size~\cite{heninger2015neighborhoods,heninger2018perturbation}. In the above 2-d example, the $x-$coordinate is used for the Koopman analysis since the stretching and folding mechanism is most clear in this direction. In a general situation, we have no idea whatsoever which are important directions. It would be good if a systematic procedure is developed for this purpose.  In practice, very often, only partial data is collected for a system and the desired one may not even be included. The problem of how much and what kind of information could be extracted from such data seems to be of practical importance.

\begin{acknowledgments}
	
This work was supported by the National Natural Science Foundation of China under Grants No. 11775035 and No. 11375093, and also by the Fundamental Research Funds for the Central Universities with Contract No. 2019XD-A10.

\end{acknowledgments}

\bibliographystyle{apsrev4-1}


\begin{thebibliography}{40}%
	\makeatletter
	\providecommand \@ifxundefined [1]{%
		\@ifx{#1\undefined}
	}%
	\providecommand \@ifnum [1]{%
		\ifnum #1\expandafter \@firstoftwo
		\else \expandafter \@secondoftwo
		\fi
	}%
	\providecommand \@ifx [1]{%
		\ifx #1\expandafter \@firstoftwo
		\else \expandafter \@secondoftwo
		\fi
	}%
	\providecommand \natexlab [1]{#1}%
	\providecommand \enquote  [1]{``#1''}%
	\providecommand \bibnamefont  [1]{#1}%
	\providecommand \bibfnamefont [1]{#1}%
	\providecommand \citenamefont [1]{#1}%
	\providecommand \href@noop [0]{\@secondoftwo}%
	\providecommand \href [0]{\begingroup \@sanitize@url \@href}%
	\providecommand \@href[1]{\@@startlink{#1}\@@href}%
	\providecommand \@@href[1]{\endgroup#1\@@endlink}%
	\providecommand \@sanitize@url [0]{\catcode `\\12\catcode `\$12\catcode
		`\&12\catcode `\#12\catcode `\^12\catcode `\_12\catcode `\%12\relax}%
	\providecommand \@@startlink[1]{}%
	\providecommand \@@endlink[0]{}%
	\providecommand \url  [0]{\begingroup\@sanitize@url \@url }%
	\providecommand \@url [1]{\endgroup\@href {#1}{\urlprefix }}%
	\providecommand \urlprefix  [0]{URL }%
	\providecommand \Eprint [0]{\href }%
	\providecommand \doibase [0]{https://doi.org/}%
	\providecommand \selectlanguage [0]{\@gobble}%
	\providecommand \bibinfo  [0]{\@secondoftwo}%
	\providecommand \bibfield  [0]{\@secondoftwo}%
	\providecommand \translation [1]{[#1]}%
	\providecommand \BibitemOpen [0]{}%
	\providecommand \bibitemStop [0]{}%
	\providecommand \bibitemNoStop [0]{.\EOS\space}%
	\providecommand \EOS [0]{\spacefactor3000\relax}%
	\providecommand \BibitemShut  [1]{\csname bibitem#1\endcsname}%
	\let\auto@bib@innerbib\@empty
	\bibitem [{\citenamefont {Hassoun}\ \emph {et~al.}(1995)\citenamefont {Hassoun}
		\emph {et~al.}}]{hassoun1995fundamentals}%
	\BibitemOpen
	\bibfield  {author} {\bibinfo {author} {\bibfnamefont {M.~H.}\ \bibnamefont
			{Hassoun}} \emph {et~al.},\ }\href@noop {} {\emph {\bibinfo {title}
			{Fundamentals of artificial neural networks}}}\ (\bibinfo  {publisher} {MIT
		press},\ \bibinfo {year} {1995})\BibitemShut {NoStop}%
	\bibitem [{\citenamefont {Strogatz}(2001)}]{strogatz2001nonlinear}%
	\BibitemOpen
	\bibfield  {author} {\bibinfo {author} {\bibfnamefont {S.}~\bibnamefont
			{Strogatz}},\ }\href@noop {} {\emph {\bibinfo {title} {Nonlinear dynamics and
				chaos: with applications to physics, biology, chemistry, and engineering
				(studies in nonlinearity)}}}\ (\bibinfo  {publisher} {CRC Press},\ \bibinfo
	{year} {2001})\BibitemShut {NoStop}%
	\bibitem [{\citenamefont {Holmes}\ \emph {et~al.}(1996)\citenamefont {Holmes},
		\citenamefont {Lumley},\ and\ \citenamefont {Berkooz}}]{holmes1996coherent}%
	\BibitemOpen
	\bibfield  {author} {\bibinfo {author} {\bibfnamefont {P.}~\bibnamefont
			{Holmes}}, \bibinfo {author} {\bibfnamefont {J.~L.}\ \bibnamefont {Lumley}},\
		and\ \bibinfo {author} {\bibfnamefont {G.}~\bibnamefont {Berkooz}},\
	}\href@noop {} {\emph {\bibinfo {title} {Turbulence, Coherent Structures,
				Dynamical Systems and Symmetry.}}}\ (\bibinfo  {publisher} {Cambridge
		University Press},\ \bibinfo {year} {1996})\BibitemShut {NoStop}%
	\bibitem [{\citenamefont {Landau}(1973)}]{landau1973statistical}%
	\BibitemOpen
	\bibfield  {author} {\bibinfo {author} {\bibfnamefont {L.}~\bibnamefont
			{Landau}},\ }\href@noop {} {\emph {\bibinfo {title} {Statistical
				Mechanics}}}\ (\bibinfo  {publisher} {Butterworth-Heinemann},\ \bibinfo
	{year} {1973})\BibitemShut {NoStop}%
	\bibitem [{\citenamefont {Prigogine}(2017)}]{prigogine2017non}%
	\BibitemOpen
	\bibfield  {author} {\bibinfo {author} {\bibfnamefont {I.}~\bibnamefont
			{Prigogine}},\ }\href@noop {} {\emph {\bibinfo {title} {Non-equilibrium
				statistical mechanics}}}\ (\bibinfo  {publisher} {Courier Dover
		Publications},\ \bibinfo {year} {2017})\BibitemShut {NoStop}%
	\bibitem [{\citenamefont {Koopman}(1931)}]{koopman1931hamiltonian}%
	\BibitemOpen
	\bibfield  {author} {\bibinfo {author} {\bibfnamefont {B.~O.}\ \bibnamefont
			{Koopman}},\ }\href@noop {} {\bibfield  {journal} {\bibinfo  {journal} {Proc.
				Natl. Acad. Sci. U.S.A.}\ }\textbf {\bibinfo {volume} {17}},\ \bibinfo
		{pages} {315} (\bibinfo {year} {1931})}\BibitemShut {NoStop}%
	\bibitem [{\citenamefont {Gaspard}\ \emph {et~al.}(1995)\citenamefont
		{Gaspard}, \citenamefont {Nicolis}, \citenamefont {Provata},\ and\
		\citenamefont {Tasaki}}]{gaspard1995spectral}%
	\BibitemOpen
	\bibfield  {author} {\bibinfo {author} {\bibfnamefont {P.}~\bibnamefont
			{Gaspard}}, \bibinfo {author} {\bibfnamefont {G.}~\bibnamefont {Nicolis}},
		\bibinfo {author} {\bibfnamefont {A.}~\bibnamefont {Provata}},\ and\ \bibinfo
		{author} {\bibfnamefont {S.}~\bibnamefont {Tasaki}},\ }\href@noop {}
	{\bibfield  {journal} {\bibinfo  {journal} {Phys. Rev. E}\ }\textbf {\bibinfo
			{volume} {51}},\ \bibinfo {pages} {74} (\bibinfo {year} {1995})}\BibitemShut
	{NoStop}%
	\bibitem [{\citenamefont {Gaspard}\ and\ \citenamefont
		{Tasaki}(2001)}]{gaspard2001liouvillian}%
	\BibitemOpen
	\bibfield  {author} {\bibinfo {author} {\bibfnamefont {P.}~\bibnamefont
			{Gaspard}}\ and\ \bibinfo {author} {\bibfnamefont {S.}~\bibnamefont
			{Tasaki}},\ }\href@noop {} {\bibfield  {journal} {\bibinfo  {journal} {Phys.
				Rev. E}\ }\textbf {\bibinfo {volume} {64}},\ \bibinfo {pages} {056232}
		(\bibinfo {year} {2001})}\BibitemShut {NoStop}%
	\bibitem [{\citenamefont {Budi{\v{s}}i{\'c}}\ \emph {et~al.}(2012)\citenamefont
		{Budi{\v{s}}i{\'c}}, \citenamefont {Mohr},\ and\ \citenamefont
		{Mezi{\'c}}}]{budivsic2012applied}%
	\BibitemOpen
	\bibfield  {author} {\bibinfo {author} {\bibfnamefont {M.}~\bibnamefont
			{Budi{\v{s}}i{\'c}}}, \bibinfo {author} {\bibfnamefont {R.}~\bibnamefont
			{Mohr}},\ and\ \bibinfo {author} {\bibfnamefont {I.}~\bibnamefont
			{Mezi{\'c}}},\ }\href@noop {} {\bibfield  {journal} {\bibinfo  {journal}
			{Chaos}\ }\textbf {\bibinfo {volume} {22}},\ \bibinfo {pages} {047510}
		(\bibinfo {year} {2012})}\BibitemShut {NoStop}%
	\bibitem [{\citenamefont {Mezi{\'c}}\ and\ \citenamefont
		{Banaszuk}(2004)}]{mezic2004comparison}%
	\BibitemOpen
	\bibfield  {author} {\bibinfo {author} {\bibfnamefont {I.}~\bibnamefont
			{Mezi{\'c}}}\ and\ \bibinfo {author} {\bibfnamefont {A.}~\bibnamefont
			{Banaszuk}},\ }\href@noop {} {\bibfield  {journal} {\bibinfo  {journal}
			{Physica D}\ }\textbf {\bibinfo {volume} {197}},\ \bibinfo {pages} {101}
		(\bibinfo {year} {2004})}\BibitemShut {NoStop}%
	\bibitem [{\citenamefont {Mezi{\'c}}(2005)}]{mezic2005spectral}%
	\BibitemOpen
	\bibfield  {author} {\bibinfo {author} {\bibfnamefont {I.}~\bibnamefont
			{Mezi{\'c}}},\ }\href@noop {} {\bibfield  {journal} {\bibinfo  {journal}
			{Nonl. Dyn.}\ }\textbf {\bibinfo {volume} {41}},\ \bibinfo {pages} {309}
		(\bibinfo {year} {2005})}\BibitemShut {NoStop}%
	\bibitem [{\citenamefont {Susuki}\ and\ \citenamefont
		{Mezi{\'c}}(2011)}]{susuki2011nonlinear}%
	\BibitemOpen
	\bibfield  {author} {\bibinfo {author} {\bibfnamefont {Y.}~\bibnamefont
			{Susuki}}\ and\ \bibinfo {author} {\bibfnamefont {I.}~\bibnamefont
			{Mezi{\'c}}},\ }\href@noop {} {\bibfield  {journal} {\bibinfo  {journal}
			{IEEE Trans. Power Syst.}\ }\textbf {\bibinfo {volume} {26}},\ \bibinfo
		{pages} {1894} (\bibinfo {year} {2011})}\BibitemShut {NoStop}%
	\bibitem [{\citenamefont {Susuki}\ and\ \citenamefont
		{Mezi{\'c}}(2012)}]{susuki2012nonlinear}%
	\BibitemOpen
	\bibfield  {author} {\bibinfo {author} {\bibfnamefont {Y.}~\bibnamefont
			{Susuki}}\ and\ \bibinfo {author} {\bibfnamefont {I.}~\bibnamefont
			{Mezi{\'c}}},\ }\href@noop {} {\bibfield  {journal} {\bibinfo  {journal}
			{IEEE Trans. Power Syst.}\ }\textbf {\bibinfo {volume} {27}},\ \bibinfo
		{pages} {1182} (\bibinfo {year} {2012})}\BibitemShut {NoStop}%
	\bibitem [{\citenamefont {Eisenhower}\ \emph {et~al.}(2010)\citenamefont
		{Eisenhower}, \citenamefont {Maile}, \citenamefont {Fischer},\ and\
		\citenamefont {Mezi{\'c}}}]{eisenhower2010decomposing}%
	\BibitemOpen
	\bibfield  {author} {\bibinfo {author} {\bibfnamefont {B.}~\bibnamefont
			{Eisenhower}}, \bibinfo {author} {\bibfnamefont {T.}~\bibnamefont {Maile}},
		\bibinfo {author} {\bibfnamefont {M.}~\bibnamefont {Fischer}},\ and\ \bibinfo
		{author} {\bibfnamefont {I.}~\bibnamefont {Mezi{\'c}}},\ }\href@noop {}
	{\bibfield  {journal} {\bibinfo  {journal} {Proc. SimBuild}\ }\textbf
		{\bibinfo {volume} {4}},\ \bibinfo {pages} {434} (\bibinfo {year}
		{2010})}\BibitemShut {NoStop}%
	\bibitem [{\citenamefont {Georgescu}\ \emph {et~al.}(2012)\citenamefont
		{Georgescu}, \citenamefont {Eisenhower},\ and\ \citenamefont
		{Mezi\'{c}}}]{georgescu2012creating}%
	\BibitemOpen
	\bibfield  {author} {\bibinfo {author} {\bibfnamefont {M.}~\bibnamefont
			{Georgescu}}, \bibinfo {author} {\bibfnamefont {B.}~\bibnamefont
			{Eisenhower}},\ and\ \bibinfo {author} {\bibfnamefont {I.}~\bibnamefont
			{Mezi\'{c}}},\ }\href@noop {} {\bibfield  {journal} {\bibinfo  {journal}
			{Proc. SimBuild}\ }\textbf {\bibinfo {volume} {5}},\ \bibinfo {pages} {40}
		(\bibinfo {year} {2012})}\BibitemShut {NoStop}%
	\bibitem [{\citenamefont {Schmid}\ \emph {et~al.}(2011)\citenamefont {Schmid},
		\citenamefont {Li}, \citenamefont {Juniper},\ and\ \citenamefont
		{Pust}}]{schmid2011applications}%
	\BibitemOpen
	\bibfield  {author} {\bibinfo {author} {\bibfnamefont {P.~J.}\ \bibnamefont
			{Schmid}}, \bibinfo {author} {\bibfnamefont {L.}~\bibnamefont {Li}}, \bibinfo
		{author} {\bibfnamefont {M.~P.}\ \bibnamefont {Juniper}},\ and\ \bibinfo
		{author} {\bibfnamefont {O.}~\bibnamefont {Pust}},\ }\href@noop {} {\bibfield
		{journal} {\bibinfo  {journal} {Theor. Comput. Fluid Dyn.}\ }\textbf
		{\bibinfo {volume} {25}},\ \bibinfo {pages} {249} (\bibinfo {year}
		{2011})}\BibitemShut {NoStop}%
	\bibitem [{\citenamefont {Bagheri}(2013)}]{bagheri2013koopman}%
	\BibitemOpen
	\bibfield  {author} {\bibinfo {author} {\bibfnamefont {S.}~\bibnamefont
			{Bagheri}},\ }\href@noop {} {\bibfield  {journal} {\bibinfo  {journal} {J.
				Fluid Mech.}\ }\textbf {\bibinfo {volume} {726}},\ \bibinfo {pages} {596}
		(\bibinfo {year} {2013})}\BibitemShut {NoStop}%
	\bibitem [{\citenamefont {Mezi{\'c}}(2013)}]{mezic2013analysis}%
	\BibitemOpen
	\bibfield  {author} {\bibinfo {author} {\bibfnamefont {I.}~\bibnamefont
			{Mezi{\'c}}},\ }\href@noop {} {\bibfield  {journal} {\bibinfo  {journal}
			{Annu. Rev. Fluid Mech.}\ }\textbf {\bibinfo {volume} {45}},\ \bibinfo
		{pages} {357} (\bibinfo {year} {2013})}\BibitemShut {NoStop}%
	\bibitem [{\citenamefont {Schmid}(2010)}]{schmid2010dynamic}%
	\BibitemOpen
	\bibfield  {author} {\bibinfo {author} {\bibfnamefont {P.~J.}\ \bibnamefont
			{Schmid}},\ }\href@noop {} {\bibfield  {journal} {\bibinfo  {journal} {J.
				Fluid Mech.}\ }\textbf {\bibinfo {volume} {656}},\ \bibinfo {pages} {5}
		(\bibinfo {year} {2010})}\BibitemShut {NoStop}%
	\bibitem [{\citenamefont {Korda}\ \emph {et~al.}(2020)\citenamefont {Korda},
		\citenamefont {Putinar},\ and\ \citenamefont {Mezi{\'c}}}]{korda2020data}%
	\BibitemOpen
	\bibfield  {author} {\bibinfo {author} {\bibfnamefont {M.}~\bibnamefont
			{Korda}}, \bibinfo {author} {\bibfnamefont {M.}~\bibnamefont {Putinar}},\
		and\ \bibinfo {author} {\bibfnamefont {I.}~\bibnamefont {Mezi{\'c}}},\
	}\href@noop {} {\bibfield  {journal} {\bibinfo  {journal} {Appl. Comput.
				Harmon. Anal.}\ }\textbf {\bibinfo {volume} {48}},\ \bibinfo {pages} {599}
		(\bibinfo {year} {2020})}\BibitemShut {NoStop}%
	\bibitem [{\citenamefont {Arbabi}\ and\ \citenamefont
		{Mezic}(2017)}]{arbabi2017ergodic}%
	\BibitemOpen
	\bibfield  {author} {\bibinfo {author} {\bibfnamefont {H.}~\bibnamefont
			{Arbabi}}\ and\ \bibinfo {author} {\bibfnamefont {I.}~\bibnamefont {Mezic}},\
	}\href@noop {} {\bibfield  {journal} {\bibinfo  {journal} {SIAM J. Appl. Dyn.
				Syst.}\ }\textbf {\bibinfo {volume} {16}},\ \bibinfo {pages} {2096} (\bibinfo
		{year} {2017})}\BibitemShut {NoStop}%
	\bibitem [{\citenamefont {Lan}\ and\ \citenamefont
		{Mezi{\'c}}(2013)}]{lan2013linearization}%
	\BibitemOpen
	\bibfield  {author} {\bibinfo {author} {\bibfnamefont {Y.}~\bibnamefont
			{Lan}}\ and\ \bibinfo {author} {\bibfnamefont {I.}~\bibnamefont
			{Mezi{\'c}}},\ }\href@noop {} {\bibfield  {journal} {\bibinfo  {journal}
			{Physica D}\ }\textbf {\bibinfo {volume} {242}},\ \bibinfo {pages} {42}
		(\bibinfo {year} {2013})}\BibitemShut {NoStop}%
	\bibitem [{\citenamefont {Morse}\ and\ \citenamefont
		{Hedlund}(1938)}]{morse1938symbolic}%
	\BibitemOpen
	\bibfield  {author} {\bibinfo {author} {\bibfnamefont {M.}~\bibnamefont
			{Morse}}\ and\ \bibinfo {author} {\bibfnamefont {G.~A.}\ \bibnamefont
			{Hedlund}},\ }\href@noop {} {\bibfield  {journal} {\bibinfo  {journal} {Am.
				J. Math.}\ }\textbf {\bibinfo {volume} {60}},\ \bibinfo {pages} {815}
		(\bibinfo {year} {1938})}\BibitemShut {NoStop}%
	\bibitem [{\citenamefont {Crutchfield}\ and\ \citenamefont
		{Packard}(1982)}]{crutchfield1982symbolic}%
	\BibitemOpen
	\bibfield  {author} {\bibinfo {author} {\bibfnamefont {J.}~\bibnamefont
			{Crutchfield}}\ and\ \bibinfo {author} {\bibfnamefont {N.}~\bibnamefont
			{Packard}},\ }\href@noop {} {\bibfield  {journal} {\bibinfo  {journal} {Int.
				J. Theor. Phys.}\ }\textbf {\bibinfo {volume} {21}},\ \bibinfo {pages} {433}
		(\bibinfo {year} {1982})}\BibitemShut {NoStop}%
	\bibitem [{\citenamefont {Robinson}(1998)}]{robinson1998dynamical}%
	\BibitemOpen
	\bibfield  {author} {\bibinfo {author} {\bibfnamefont {C.}~\bibnamefont
			{Robinson}},\ }\href@noop {} {\emph {\bibinfo {title} {Dynamical systems:
				stability, symbolic dynamics, and chaos}}}\ (\bibinfo  {publisher} {CRC
		press},\ \bibinfo {year} {1998})\BibitemShut {NoStop}%
	\bibitem [{\citenamefont {Hao}(1991)}]{hao1991symbolic}%
	\BibitemOpen
	\bibfield  {author} {\bibinfo {author} {\bibfnamefont {B.-l.}\ \bibnamefont
			{Hao}},\ }\href@noop {} {\bibfield  {journal} {\bibinfo  {journal} {Physica
				D}\ }\textbf {\bibinfo {volume} {51}},\ \bibinfo {pages} {161} (\bibinfo
		{year} {1991})}\BibitemShut {NoStop}%
	\bibitem [{\citenamefont {Biham}\ and\ \citenamefont
		{Wenzel}(1989)}]{biham1989characterization}%
	\BibitemOpen
	\bibfield  {author} {\bibinfo {author} {\bibfnamefont {O.}~\bibnamefont
			{Biham}}\ and\ \bibinfo {author} {\bibfnamefont {W.}~\bibnamefont {Wenzel}},\
	}\href@noop {} {\bibfield  {journal} {\bibinfo  {journal} {Phys. Rev. Lett.}\
		}\textbf {\bibinfo {volume} {63}},\ \bibinfo {pages} {819} (\bibinfo {year}
		{1989})}\BibitemShut {NoStop}%
	\bibitem [{\citenamefont {Jaeger}\ and\ \citenamefont
		{Kantz}(1997)}]{jaeger1997structure}%
	\BibitemOpen
	\bibfield  {author} {\bibinfo {author} {\bibfnamefont {L.}~\bibnamefont
			{Jaeger}}\ and\ \bibinfo {author} {\bibfnamefont {H.}~\bibnamefont {Kantz}},\
	}\href@noop {} {\bibfield  {journal} {\bibinfo  {journal} {J. Phys. A: Math.
				Gen.}\ }\textbf {\bibinfo {volume} {30}},\ \bibinfo {pages} {L567} (\bibinfo
		{year} {1997})}\BibitemShut {NoStop}%
	\bibitem [{\citenamefont {Grassberger}\ and\ \citenamefont
		{Kantz}(1985)}]{grassberger1985generating}%
	\BibitemOpen
	\bibfield  {author} {\bibinfo {author} {\bibfnamefont {P.}~\bibnamefont
			{Grassberger}}\ and\ \bibinfo {author} {\bibfnamefont {H.}~\bibnamefont
			{Kantz}},\ }\href@noop {} {\bibfield  {journal} {\bibinfo  {journal} {Phys.
				Lett. A}\ }\textbf {\bibinfo {volume} {113}},\ \bibinfo {pages} {235}
		(\bibinfo {year} {1985})}\BibitemShut {NoStop}%
	\bibitem [{\citenamefont {Alla}\ and\ \citenamefont
		{Kutz}(2017)}]{alla2017nonlinear}%
	\BibitemOpen
	\bibfield  {author} {\bibinfo {author} {\bibfnamefont {A.}~\bibnamefont
			{Alla}}\ and\ \bibinfo {author} {\bibfnamefont {J.~N.}\ \bibnamefont
			{Kutz}},\ }\href@noop {} {\bibfield  {journal} {\bibinfo  {journal} {SIAM J.
				Sci. Comput.}\ }\textbf {\bibinfo {volume} {39}},\ \bibinfo {pages} {B778}
		(\bibinfo {year} {2017})}\BibitemShut {NoStop}%
	\bibitem [{\citenamefont {Brunton}\ \emph {et~al.}(2016)\citenamefont
		{Brunton}, \citenamefont {Brunton}, \citenamefont {Proctor},\ and\
		\citenamefont {Kutz}}]{brunton2016koopman}%
	\BibitemOpen
	\bibfield  {author} {\bibinfo {author} {\bibfnamefont {S.~L.}\ \bibnamefont
			{Brunton}}, \bibinfo {author} {\bibfnamefont {B.~W.}\ \bibnamefont
			{Brunton}}, \bibinfo {author} {\bibfnamefont {J.~L.}\ \bibnamefont
			{Proctor}},\ and\ \bibinfo {author} {\bibfnamefont {J.~N.}\ \bibnamefont
			{Kutz}},\ }\href@noop {} {\bibfield  {journal} {\bibinfo  {journal} {PLoS
				One}\ }\textbf {\bibinfo {volume} {11}} (\bibinfo {year} {2016})}\BibitemShut
	{NoStop}%
	\bibitem [{\citenamefont {Cvitanovi\'c}\ \emph {et~al.}(2020)\citenamefont
		{Cvitanovi\'c}, \citenamefont {Artuso}, \citenamefont {Mainieri},
		\citenamefont {Tanner},\ and\ \citenamefont {Vattay}}]{cvitanovic2020chaos}%
	\BibitemOpen
	\bibfield  {author} {\bibinfo {author} {\bibfnamefont {P.}~\bibnamefont
			{Cvitanovi\'c}}, \bibinfo {author} {\bibfnamefont {R.}~\bibnamefont
			{Artuso}}, \bibinfo {author} {\bibfnamefont {R.}~\bibnamefont {Mainieri}},
		\bibinfo {author} {\bibfnamefont {G.}~\bibnamefont {Tanner}},\ and\ \bibinfo
		{author} {\bibfnamefont {G.}~\bibnamefont {Vattay}},\ }\href@noop {} {\emph
		{\bibinfo {title} {Chaos: Classical and Quantum}}}\ (\bibinfo  {publisher}
	{ChaosBook.org},\ \bibinfo {year} {2020})\BibitemShut {NoStop}%
	\bibitem [{\citenamefont {Govindarajan}\ \emph {et~al.}(2019)\citenamefont
		{Govindarajan}, \citenamefont {Mohr}, \citenamefont {Chandrasekaran},\ and\
		\citenamefont {Mezic}}]{govindarajan2019approximation}%
	\BibitemOpen
	\bibfield  {author} {\bibinfo {author} {\bibfnamefont {N.}~\bibnamefont
			{Govindarajan}}, \bibinfo {author} {\bibfnamefont {R.}~\bibnamefont {Mohr}},
		\bibinfo {author} {\bibfnamefont {S.}~\bibnamefont {Chandrasekaran}},\ and\
		\bibinfo {author} {\bibfnamefont {I.}~\bibnamefont {Mezic}},\ }\href@noop {}
	{\bibfield  {journal} {\bibinfo  {journal} {SIAM J. Appl. Dyn. Syst.}\
		}\textbf {\bibinfo {volume} {18}},\ \bibinfo {pages} {1454} (\bibinfo {year}
		{2019})}\BibitemShut {NoStop}%
	\bibitem [{\citenamefont {Brunton}\ \emph {et~al.}(2017)\citenamefont
		{Brunton}, \citenamefont {Brunton}, \citenamefont {Proctor}, \citenamefont
		{Kaiser},\ and\ \citenamefont {Kutz}}]{brunton2017chaos}%
	\BibitemOpen
	\bibfield  {author} {\bibinfo {author} {\bibfnamefont {S.~L.}\ \bibnamefont
			{Brunton}}, \bibinfo {author} {\bibfnamefont {B.~W.}\ \bibnamefont
			{Brunton}}, \bibinfo {author} {\bibfnamefont {J.~L.}\ \bibnamefont
			{Proctor}}, \bibinfo {author} {\bibfnamefont {E.}~\bibnamefont {Kaiser}},\
		and\ \bibinfo {author} {\bibfnamefont {J.~N.}\ \bibnamefont {Kutz}},\
	}\href@noop {} {\bibfield  {journal} {\bibinfo  {journal} {Nat. Commun.}\
		}\textbf {\bibinfo {volume} {8}},\ \bibinfo {pages} {1} (\bibinfo {year}
		{2017})}\BibitemShut {NoStop}%
	\bibitem [{\citenamefont {Heninger}\ \emph {et~al.}(2015)\citenamefont
		{Heninger}, \citenamefont {Lippolis},\ and\ \citenamefont
		{Cvitanovi{\'c}}}]{heninger2015neighborhoods}%
	\BibitemOpen
	\bibfield  {author} {\bibinfo {author} {\bibfnamefont {J.~M.}\ \bibnamefont
			{Heninger}}, \bibinfo {author} {\bibfnamefont {D.}~\bibnamefont {Lippolis}},\
		and\ \bibinfo {author} {\bibfnamefont {P.}~\bibnamefont {Cvitanovi{\'c}}},\
	}\href@noop {} {\bibfield  {journal} {\bibinfo  {journal} {Phys. Rev. E}\
		}\textbf {\bibinfo {volume} {92}},\ \bibinfo {pages} {062922} (\bibinfo
		{year} {2015})}\BibitemShut {NoStop}%
	\bibitem [{\citenamefont {Heninger}\ \emph {et~al.}(2018)\citenamefont
		{Heninger}, \citenamefont {Lippolis},\ and\ \citenamefont
		{Cvitanovi{\'c}}}]{heninger2018perturbation}%
	\BibitemOpen
	\bibfield  {author} {\bibinfo {author} {\bibfnamefont {J.~M.}\ \bibnamefont
			{Heninger}}, \bibinfo {author} {\bibfnamefont {D.}~\bibnamefont {Lippolis}},\
		and\ \bibinfo {author} {\bibfnamefont {P.}~\bibnamefont {Cvitanovi{\'c}}},\
	}\href@noop {} {\bibfield  {journal} {\bibinfo  {journal} {Commun. Nonl. Sci.
				Numer. Simul.}\ }\textbf {\bibinfo {volume} {55}},\ \bibinfo {pages} {16}
		(\bibinfo {year} {2018})}\BibitemShut {NoStop}%
	\bibitem [{\citenamefont {Sim{\'o}}(1979)}]{simo1979henon}%
	\BibitemOpen
	\bibfield  {author} {\bibinfo {author} {\bibfnamefont {C.}~\bibnamefont
			{Sim{\'o}}},\ }\href@noop {} {\bibfield  {journal} {\bibinfo  {journal} {J.
				Stat. Phys.}\ }\textbf {\bibinfo {volume} {21}},\ \bibinfo {pages} {465}
		(\bibinfo {year} {1979})}\BibitemShut {NoStop}%
	\bibitem [{\citenamefont {Gonchenko}\ and\ \citenamefont
		{Gonchenko}(2016)}]{gonchenko2016variety}%
	\BibitemOpen
	\bibfield  {author} {\bibinfo {author} {\bibfnamefont {A.}~\bibnamefont
			{Gonchenko}}\ and\ \bibinfo {author} {\bibfnamefont {S.}~\bibnamefont
			{Gonchenko}},\ }\href@noop {} {\bibfield  {journal} {\bibinfo  {journal}
			{Physica D}\ }\textbf {\bibinfo {volume} {337}},\ \bibinfo {pages} {43}
		(\bibinfo {year} {2016})}\BibitemShut {NoStop}%
	\bibitem [{\citenamefont {Gonchenko}\ \emph {et~al.}(2014)\citenamefont
		{Gonchenko}, \citenamefont {Gonchenko}, \citenamefont {Kazakov},\ and\
		\citenamefont {Turaev}}]{gonchenko2014simple}%
	\BibitemOpen
	\bibfield  {author} {\bibinfo {author} {\bibfnamefont {A.}~\bibnamefont
			{Gonchenko}}, \bibinfo {author} {\bibfnamefont {S.}~\bibnamefont
			{Gonchenko}}, \bibinfo {author} {\bibfnamefont {A.}~\bibnamefont {Kazakov}},\
		and\ \bibinfo {author} {\bibfnamefont {D.}~\bibnamefont {Turaev}},\
	}\href@noop {} {\bibfield  {journal} {\bibinfo  {journal} {Int. J. Bifurc.
				Chaos}\ }\textbf {\bibinfo {volume} {24}},\ \bibinfo {pages} {1440005}
		(\bibinfo {year} {2014})}\BibitemShut {NoStop}%
	\bibitem [{\citenamefont {Gonchenko}\ \emph {et~al.}(2005)\citenamefont
		{Gonchenko}, \citenamefont {Ovsyannikov}, \citenamefont {Sim{\'o}},\ and\
		\citenamefont {Turaev}}]{gonchenko2005three}%
	\BibitemOpen
	\bibfield  {author} {\bibinfo {author} {\bibfnamefont {S.~V.}\ \bibnamefont
			{Gonchenko}}, \bibinfo {author} {\bibfnamefont {I.}~\bibnamefont
			{Ovsyannikov}}, \bibinfo {author} {\bibfnamefont {C.}~\bibnamefont
			{Sim{\'o}}},\ and\ \bibinfo {author} {\bibfnamefont {D.}~\bibnamefont
			{Turaev}},\ }\href@noop {} {\bibfield  {journal} {\bibinfo  {journal} {Int.
				J. Bifurc. Chaos}\ }\textbf {\bibinfo {volume} {15}},\ \bibinfo {pages}
		{3493} (\bibinfo {year} {2005})}\BibitemShut {NoStop}%
\end{thebibliography}
%

\end{document}